\documentclass[aps, preprint, supberscriptaddress, showpacs, preprintnumbers, amsmath, amssymb, nofootinbib, longbibliography]{revtex4-1}

\usepackage{graphicx}
\usepackage{braket}
\usepackage[
 	colorlinks=true, 
   pdfstartview=FitV, 
    linkcolor=blue, 
    citecolor=magenta, 
    urlcolor=blue, 
    bookmarks=true,
    bookmarksnumbered=true,
    pdftitle={},
    pdfauthor={Shoichiro Tsutsui and Takahiro M. Doi}
    ]{hyperref}

\newcommand{\fig}[2]{\includegraphics[width=#1]{#2}}
\renewcommand{\Re}{\mathrm{Re}}
\renewcommand{\Im}{\mathrm{Im}}
\newcommand{\del}{\partial}

\newcommand{\complex}{\mathbb{C}}
\newcommand{\integer}{\mathbb{Z}}
\newcommand{\calD}{\mathcal{D}}
\newcommand{\calJ}{\mathcal{J}}
\newcommand{\calK}{\mathcal{K}}

\newcommand{\lp}{\left(}
\newcommand{\rp}{\right)}

\begin{document}

\title{The title}
\title{Distribution of solutions of the fastest apparent convergence condition in optimized perturbation theory and its relation to anti-Stokes lines}

\author{Shoichiro Tsutsui}
\email[]{stsutsui@post.kek.jp}
\affiliation{KEK, High Energy Accelerator Research Organization,1-1 Oho, Tsukuba, Ibaraki, 305-0801, Japan}

\author{Takahiro M. Doi}
\affiliation{Theoretical Research Division, Nishina Center, RIKEN, Wako, 351-0198, Japan}

\begin{abstract}%
We discuss fundamental properties of the fastest apparent convergence (FAC) condition which is used as a variational criterion in optimized perturbation theory (OPT).
We examine an integral representation of the FAC condition and a distribution of the zeros of the integral in a complex artificial parameter space on the basis of theory of Lefschetz thimbles.
We find that the zeros accumulate on a certain line segment so-called anti-Stokes line in the limit $K \to \infty$, where $K$ is a truncation order of a perturbation series.
This phenomenon gives an underlying mechanism that physical quantities calculated by OPT can be insensitive to the
choice of the artificial parameter.
\end{abstract}

\maketitle

\section{Introduction}
Perturbation theory would be the first option to perform analytic calculation in many fields of physics.
However, the usual perturbation theory cannot capture nonperturbative nature of models because a perturbation series is not a convergent series but just an asymptotic series.
To address the problem of the perturbation theory, various methods to improve the convergence properties of the perturbation series are suggested.

Optimized perturbation theory (OPT) considered in this paper is one of a resummation technique first developed for quantum mechanics~\cite{Caswell:1979qh,Seznec:1979ev,Halliday:1979vn,Killingbeck1981} and immediately extended to apply quantum field theories~\cite{Stevenson:1981vj,Stevenson:1982qw,Okopinska:1987hp,Stancu:1989sk,Chiku:1998kd}. (See also~\cite{Arteca, Kleinert,Andersen:2004fp,Jakovac} and references therein).

The basic idea of OPT is combining the usual perturbation theory and variational methods.
In OPT, one introduces an artificial parameter, say $z$, and reorganizes an interaction term of an action.
After performing the perturbative expansion in terms of the modified interaction term, one obtains physical quantities depending on the artificial parameter.
When one imposes an appropriate variational criterion, $z$ has a nontrivial dependence in terms of an expansion parameter of the usual perturbation theory like coupling constants, and thus, non-perturbative effects could be included in the physical quantities\footnote{Precisely speaking, OPT is always applicable even if an action has no natural expansion parameters like supersymmetric Yang-Mills integral or IIB matrix model ~\cite{Sugino:2001fn,Kawai:2002jk,Kawai:2002ub}.}.
Similar techniques refered to as order dependent mapping~\cite{Seznec:1979ev}, variational perturbation~\cite{KLEINERT1993332}, self-consistent expansion~\cite{Schwartz1992}, gaussian expansion method~\cite{Kabat:1999hp, Nishimura:2001sx} and so on~\cite{Bender:1987dn, Bender:1988rq, Shaverdian:1983ay} are developed in various fields of physics.

However, OPT itself does not tell us how to impose the variational criterion.
Actually, much less is known about this issue except for a few concrete examples.
An empirical way to do is requiring that higher order terms of the modified perturbation series should have lesser contribution.
This is known as the fastest apparent convergence (FAC) condition~\cite{Seznec:1979ev,Halliday:1979vn,Halliday:1979xh}.
Seemingly, the FAC condition is natural because it is a necessary condition for the convergence of the modified perturbation series.
On the other hand, there are serious conceptual issues in a use of the FAC condition.
Because exact physical quantities are independent of the unphysical parameter $z$ by definition, approximants obtained in OPT should be insensitive to a choice of the parameter.
In other words, the approximants should be locally flat functions of $z$.
However, it is not guaranteed that a solution of the FAC condition is a interior point of the flat region.

In this paper,
we answer to this issue by studying analytic properties of the FAC condition relying on its integral representation.
We show that asymptotic properties of solutions of the FAC condition
can be clarified by studying the topological structure of Lefschetz thimbles, which are a generalization of steepest descent contours.

This paper is organized as follows. 
In section~\ref{sec:opt}, we review OPT and discuss issues of the FAC condition in detail.
In section~\ref{sec:thimble}, we develop a way to analyze properties of solutions of the FAC condition  on the basis of theory of Lefschetz thimbles.
Finally in section~\ref{sec:application}, we apply our method to a one dimensional integral and discuss an underlying mechanism that a solution of the FAC condition gives a reasonable choice of the artificial parameter of OPT.
Section~\ref{sec:summary} is devoted to summary and concluding remarks.

\section{Optimized perturbation theory}\label{sec:opt}
In this section, we explain an idea of OPT.
Let $S(x; \lambda)$ is an action which involves a coupling constant $\lambda$.
Our goal is to evaluate an integral:
\begin{align}
Z(\lambda)=\int_\calD dx e^{-S(x;\lambda)},
\label{Z}
\end{align}
where $\calD$ is an integration domain. 
In OPT, 
the action is decomposed as
\begin{align} 
S(x;\lambda) = S_0(x;z) + \delta S_I(x;\lambda,z)|_{\delta=1}, 
\label{action}
\end{align}
such that $S_0(x;z)$ becomes an action of an exactly solvable model.
Here, we introduce a complex parameter $z$, which will be determined later.
One can evaluate the value of the integral~\eqref{Z}
by performing a formal expansion in terms of $\delta$.
We denote the $K$th-order approximant of the integral by $Z_K$:
\begin{align}
Z_K(\lambda, z) &\equiv
\sum_{k=0}^K a_k(\lambda,z) \delta^k\Big|_{\delta=1}, \label{Zk}
\\
a_k(\lambda, z) &=
\frac{(-1)^k}{k!} \int_\calD dx S_I^k(x; \lambda, z) e^{-S_0(x;\,z)} .
\label{ak}
\end{align}
A peculiar feature of the approximant $Z_K$ is that it depends on the unphysical parameter $z$.
Because the original quantity we hope to obtain does not have such $z$-dependence,
it is natural to require that the approximant should be insensitive to the choice of the unphysical parameter.
Thus, a possible choice of $z$ is a solution of the following equation:
\begin{align}
\frac{\partial}{\partial z}
Z_K(\lambda, z)=0.
\label{PMS}
\end{align} 
This condition is known as the principle of minimal sensitivity (PMS)~\cite{Caswell:1979qh, Stevenson:1981vj}.

Although the physical meaning of the PMS condition is clear,
it is sometimes not suitable for analytic studies when the form of Eq.~\eqref{PMS} is involved.
In that case, 
an other possible option to determine $z$ would be the fastest apparent convergence (FAC) condition, 
which requires that the higher order terms of $\delta$ in Eq.~\eqref{Zk} should have lesser contribution than the lower-order terms.
In practice,
the simplest version of the FAC condition,
\begin{align}
Z_K(\lambda, z) - Z_{K-1}(\lambda, z) = 0
\label{FAC1}
\end{align} 
is frequently imposed.
In our notation, this condition is equivalent to
\begin{align}
a_K(\lambda, z) = 0.
\label{FAC2}
\end{align}
The FAC condition is also a reasonable criterion in the sense that it is a necessary condition for the convergence of the infinite series $\lim_{K\to\infty}Z_K$.
Moreover, the FAC condition plays an essential role in several cases.
One example is an application of OPT to quantum field theories slightly away from local-equilibrium.
There, the FAC condition naturally arises in a derivation of hydrodynamic equations~\cite{Hayata:2015lga}.

An actual precision of the approximation depends on the choice of the criterion, namely the PMS or FAC condition.
On the other hand,
both conditions have a same mechanism for taking into account nonperturbative nature of the model.
Indeed, through Eq.~\eqref{PMS} or Eq.~\eqref{FAC1},
the approximant $Z_K$ has a nontrivial $\lambda$-dependence that the usual perturbation theory never achieves.
Another important feature is that the artificial parameter also depends on $K$, a truncation order of the $\delta$-expansion.
It is known that this $K$-dependence is crucial for the convergence of OPT.
Indeed, it has been proven that the infinite series $\lim_{K\to\infty}Z_K(z(K))$ converges to the exact value $Z$ if $z$ has an appropriate $K$-dependence for
one dimensional integrals~\cite{Buckley:1992pc, Bender:1993nd, Remez:2018tle} and quantum mechanical anharmonic oscillators~\cite{Halliday:1979xh, Duncan:1992ba, Guida:1994zv, Guida:1995px, Kleinert:1995hc}.
For these known cases, the solutions of the PMS and FAC conditions reproduce an appropriate $K$-dependence.

In spite of these good properties,
the FAC condition is not fully reliable.
In particular, we raise two issues in a use of the FAC condition.
The first issue is that, 
a solution of the FAC condition is not unique in general.
For instance, 
the FAC condition for a one dimensional integral we will discuss in this paper becomes a polynomial equation whose order is $2K$.
Hence, one have to put additional criteria by hand to resolve this ambiguity.
The second issues is that the FAC condition does not guarantee the insensitivity of the approximant as a function of the unphysical parameter.
In order to address these issues, we begin our discussion with studying distribution of the zeros of $a_K(\lambda, z)$ in the complex $z$-plane.

\section{Lefschetz-thimbles}\label{sec:thimble}
In this section,
we give a way to examine zeros of $a_K(\lambda, z)$, the $K$th-order coefficient of the $\delta$-expansion.
For this purpose,
the integral representation of $a_K$ Eq.~\eqref{ak} is useful as we will see below.
Let us start with defining an effective action by
\begin{align}
I_K(x;z)\equiv S_0(x;z) - K\log S_I(x;z).
\end{align}
Here, we fix the coupling constant and omit its dependence.
From Eqs.~\eqref{ak} and~\eqref{FAC2},
the FAC condition is written by 
\begin{align}
a_K(z) = \frac{(-1)^k}{k!}\int_\calD dx e^{-I_K(x;\,z)} = 0.
\label{integral rep of FAC}
\end{align}
In order to evaluate this integral,
it is convenient to deform the integration domain on the real axis into a set of steepest descent contours in a complex plane.
Such contour is characterized by a flow whose source is a saddle point of the effective action.
Let us denote the saddle point by $\sigma_i$,
and suppose that a number of the saddle points is $N$, i.e.,
\begin{align}
\frac{\del I_K(\xi; z)}{\del \xi}\Big|_{\xi = \sigma_i} = 0,
\quad 
i = 1, \dots, N.
\end{align}
We also assume that
\begin{align}
\frac{\del^2 I_K(\xi; z)}{\del \xi^2}\Big|_{\xi = \sigma_i} \neq 0,
\end{align}
and all saddle points are isolated.
Then, the steepest descent contour is obtained by
\begin{align}
\calJ_i
=
\left\{
\xi(t) \in \complex \, \Big| \,
\frac{d \xi(t)}{d t} = +\overline{\frac{\del I_K}{\del \xi}}, \,
\xi(t=0) = \sigma_i
\right\}.
\label{J}
\end{align}
This contour and its generalization to higher dimensions is known as a Lefschetz thimble.
One can easily confirm that the real part of the effective action increases monotonically along the flow.
We also note that the imaginary part of the effective action is constant along the flow.
An endpoint of the Lefschetz thimble is a point at infinity or one of the logarithmic singularities of the effective action.

Since any well-defined integrals should be represented by a linear combination of the contours $\sum_{i=1}^N n_i\calJ_i$, Eq.~\eqref{integral rep of FAC} would be written as
\begin{align}
a_K(z)
= 
\sum_{i=1}^N n_i e^{-i\Im I_K(\sigma_i;\,z)} 
\int_{\calJ_i} d\xi e^{-\Re I_K(\xi;\,z)},
\label{thimble decomposition of FAC}
\end{align}
where $n_i$ is an unknown weight factor at this moment.
Relying on this expression,
we find that $a_K(z)$ vanishes if and only if
two or more steepest descent contours contribute to $a_K(z)$ and cancel out with each other.

In order to estimate when does the cancellation occur, 
one can use the saddle point technique.
Indeed, for a general class of models,
we will find that the saddle point approximation makes sense.
To see this, 
we suppose that the action has a following form:
\begin{align}
S(x) = \frac{\omega^2}{2}x^2 + \sum_{p=3}^q \lambda_p x^p, \quad q = 4, 6, \dots,
\label{action:general form}
\end{align}
where $\lambda_p$ is an arbitrary complex constant such that the integral~\eqref{Z} converges.
The corresponding effective action is given by
\begin{align}
I_K(x) = \frac{zx^2}{2} 
- K\log \lp -\frac{z}{2}x^2 + \sum_{p=3}^q \lambda_p x^p + \frac{\omega^2}{2}x^2 \rp.
\label{effective action:general form}
\end{align}
Replacing $z$ by $K^{1-2/q}z$ and changing the variable $x$ by $K^{1/q}x$,
the effective action reads
\begin{align}
I_K(x) = 
K \left[
\frac{zx^2}{2} 
- \log \lp -\frac{z}{2}x^2 + \lambda_q x^q + O(K^{-1/q}) \rp
\right] + \text{const.}.
\label{effective action: factorized form}
\end{align}
Therefore, if the truncation order of the $\delta$-expansion $K$ is large enough,
the integrand appeared in Eq.~\eqref{thimble decomposition of FAC} has a sharp peak around its saddle point.
In the lowest order approximation,
the FAC condition becomes
\begin{align}
\sum_{i=1}^N n_i e^{-I_K(\sigma_i;\,z)} = 0.
\label{anti Stokes line}
\end{align}
A detail derivation of this equation is given in appendix~\ref{App:saddle point}.
A set of solutions of this equation forms line segments in the complex $z$-plane, and these are known as the anti-Stokes lines.
From the above discussion, we find that solutions of the FAC condition distribute on the anti-Stokes lines.
In particular, the solutions appear exactly on the anti-Stokes line in the limit $K\to\infty$.

A remaining issue is how to compute the weight factors $\{n_i\}$.
Actually, this is a complicated task since the weight factors are governed by a global structure or topology of the Lefschetz thimbles in the complex $\xi$-plane.
Fortunately, 
a concrete way to obtain $\{n_i\}$ is well-studied for finite dimensional integrals on the basis of Picard-Lefschetz theory~\cite{Pham:1983, Howls10.2307/53139, Delabaere2002}.
Since there are already many applications of this framework to physics including field theories~\cite{Witten:2010cx, Witten:2010zr}, many reviews are available. (For instance, see~\cite{Tanizaki:2015gpl} and references therein.)
Here, we just give a sketch how to compute the weight factors $\{n_i\}$.
What plays a key role is a steepest ascent contour defined by
\begin{align}
\calK_i
=
\left\{
\xi(t) \in \complex \, \Big| \,
\frac{d \xi(t)}{d t} = -\overline{\frac{\del I_K}{\del \xi}}, \,
\xi(t=0) = \sigma_i
\right\}.
\label{K}
\end{align}
We assume that different saddle points are not connected by the flows, and introduce an orientation of $\{\calJ_i\}$ and $\{\calK_i\}$ by an appropriate manner.
By their definitions, $\calJ_i$ has an intersection point with $\calK_i$ only at $\xi = \sigma_i$.
In this case,
one can define a kind of an \textit{inner product} by
\begin{align}
\braket{\calJ_i, \calK_j} = \delta_{ij}.
\end{align}
Thus, the weight factor is given by
\begin{align}
n_i = \braket{\calD, \calK_i}.
\end{align}
This means that the weight factor is obtained as an intersection number between the original integration contour $\calD$ and a steepest ascent contour $\calK_i$. 

Since the weight factor $n_i$ is integer,
it is a discontinuous function of $z$.
Sudden change of $n_i$ at certain $z$ means that topological structure of the Lefschetz thimbles changes at that point.
A set of these points again forms line segments in the complex $z$-plane.
They are referred to as the Stokes lines~\cite{Berry_1988,Berry_1989}.
For multiple integrals, it is quite difficult to determine $n_i$, and hence, the Stokes lines.
On the other hand, 
the Stokes lines can be obtained by a simple criterion exceptionally if the complex dimension is one.
In that case, each Lefschetz thimble is uniquely labeled by the imaginary part of the effective action on the thimble.
Therefore, the Stokes line is determined by
\begin{align}
\Im I_K(\sigma_i; z) - \Im I_K(\sigma_j; z) = 2\pi n, 
\quad i \neq j, \quad n \in \integer.
\end{align}
We note that nonzero $n$ is allowed because the imaginary part of the effective action jumps by $2\pi$ across a branch cut of the logarithmic function.

\section{Application}\label{sec:application}
In this section, 
we explicitly show the relation between the anti-Stokes line and a distribution of solutions of the FAC condition for a one dimensional integral.
The simplest nontrivial example of Eq.~\eqref{action:general form} would be \begin{align}
Z(\omega^2,\lambda)
= 
\int_{-\infty}^\infty \frac{dx}{\sqrt{2\pi}} e^{-S(x;\,\omega^2,\lambda)}, \quad
S(x;\omega^2,\lambda)
=\frac{\omega^2}{2}x^2 + \frac{\lambda}{4}x^4,
\label{model}
\end{align}
where $\lambda \in \mathbb{C}$ is a complex constant which obeys ${\rm Re} \lambda > 0$.
The sign of the parameter in the quadratic term $\omega^2$ affects the asymptotic behavior of the integral.
For instance,
the analytic expressions of $Z(\omega^2,\lambda)$ for $\omega^2=1,0,-1$ are given by
\begin{align}
Z(1,\lambda)
&=
\frac{1}{2\sqrt{\pi\lambda}} 
\exp\left(\frac{1}{8\lambda}\right)
K_{1/4}\left(\frac{1}{8\lambda}\right), \\
Z(0,\lambda)
&=
\frac{\Gamma(\frac{1}{4})}{2\sqrt{\pi\lambda}}, \\
Z(-1,\lambda)
&=
\frac{1}{2}\sqrt{\frac{\pi}{2\lambda}}
\exp\left(\frac{1}{8\lambda}\right)
\left(
I_{-1/4}\left(\frac{1}{8\lambda}\right) + I_{1/4}\left(\frac{1}{8\lambda}\right)
\right),
\end{align}
where 
$I_\nu(x)$ and $K_\nu(x)$ are the modified Bessel function of the first and second kind, respectively. 
The convergence property of the $\delta$-expansion is well studied for this integral~\cite{Buckley:1992pc, Bender:1993nd, Remez:2018tle}.

\subsection{$\delta$-expansion}
Let us apply OPT to evaluate the integral~\eqref{model}.
First, we decompose the action by introducing a complex constant $z$ as
\begin{align} 
&S(x;\omega^2,\lambda)=S_0(x;z)+\delta S_I(x;\omega^2,\lambda,z)|_{\delta=1}, \\
&S_0(x;z)=\frac{z}{2}x^2, \quad
S_I(x;\omega^2,\lambda,z)=
\left( \frac{\omega^2-z}{2}x^2 + \frac{\lambda}{4}x^4 \right).
\end{align}
Here, $z$ is arbitrary as long as $\Re z > 0$.
Performing the Taylor expansion of $Z$ in terms of $\delta$ up to terms of order $K$,
we get
\begin{align}
Z_K(z)
&=
\sum_{k=0}^K a_k(\omega^2,\lambda;z) \delta^k|_{\delta=1},\\
a_k(z)&=\frac{(-1)^k}{k!}
\int_{-\infty}^\infty\frac{dx}{\sqrt{2\pi}}
\left(\frac{\omega^2-z}{2}x^2 + \frac{\lambda}{4}x^4 \right)^k
e^{-z x^2/2}.
\label{ak integral expression}
\end{align} 
By using the binomial expansion and compute the gaussian integrals term-by-term,
we reach to
\begin{align}
a_k(z)=
\sqrt{\frac{\pi}{z}}
\frac{(-\lambda)^k}{k!z^{2k}}
\frac{1}{\Gamma(\frac{1}{2}-2k)}
{}_1F_1\left(-k,\frac{1}{2}-2k;\frac{(\omega^2-z)z}{\lambda}\right) 
\label{ak analytic expression},
\end{align} 
where ${}_1F_1(a,b;x)$ is the confluent hypergeometric function.
Since ${}_1F_1(a,b;x)/\Gamma(b)$ is an entire function,
$a_k(z)$ is analytic when $z\neq0$.
In particular, $z^{2k+1/2}a_k(z)$ is a polynomial function of $z$ whose order is $2k$.
Due to the symmetry under the exchange of $z \leftrightarrow (\omega^2-z)$,
the FAC condition $a_K(z) = 0$ has $K$ solutions in the right half-plane.
Numerically, they are easily obtained on the basis of the Durand-Kerner-Aberth method, for instance.
However, 
the analytic expression~\eqref{ak analytic expression} is useless to argue its distribution of zeros.
Thus, we should rely on the integral expression~\eqref{ak integral expression} rather than Eq.~\eqref{ak analytic expression} to say something about properties of the FAC condition.

\subsection{Stokes and anti-Stokes lines}
As we discuss in Sec.~\ref{sec:thimble},
we introduce the effective action $I_K$ as follows:
\begin{align}
a_K(z)
&=
\frac{(-1)^k}{\sqrt{2\pi}k!}
\int_\calD d\xi
e^{-I_K(\xi;\,z)} \label{aK quartic} \\
I_K(\xi;z)
&=
\frac{z}{2}\xi^2
-K\log\left(\frac{\omega^2-z}{2}\xi^2 + \frac{\lambda}{4}\xi^4\right),
\label{effective action: FAC}
\end{align} 
where $\calD$ is the real axis at this moment.
Thanks to the symmetry $I_K(\xi;z)=I_K(-\xi;z)$, 
we consider only the right half part of the complex $\xi$-plane.
Inside the region,
there are two saddle points, which are given by
\begin{align}
\sigma_\pm
= 
\sqrt{\frac{2K}{z}(1 + w \pm \sqrt{1 + w^2})}.
\label{saddle}
\end{align}
Here, 
we define
\begin{align}
w = \frac{z(z-\omega^2)}{2K\lambda},
\label{z to w}
\end{align}
for later convenience.
Thus, we find that the integral~\eqref{aK quartic} is rewritten in terms of integrals on the Lefschetz thimbles, which are obtained by solving the flow equation defined in Eq.~\eqref{J} from the saddle points $\sigma_\pm$.
We denote the Lefschetz thimbles by $\calJ_\pm$ and corresponding weight factors by $n_\pm$.

\begin{figure}[tbh]
	\centering
	\fig{7.5cm}{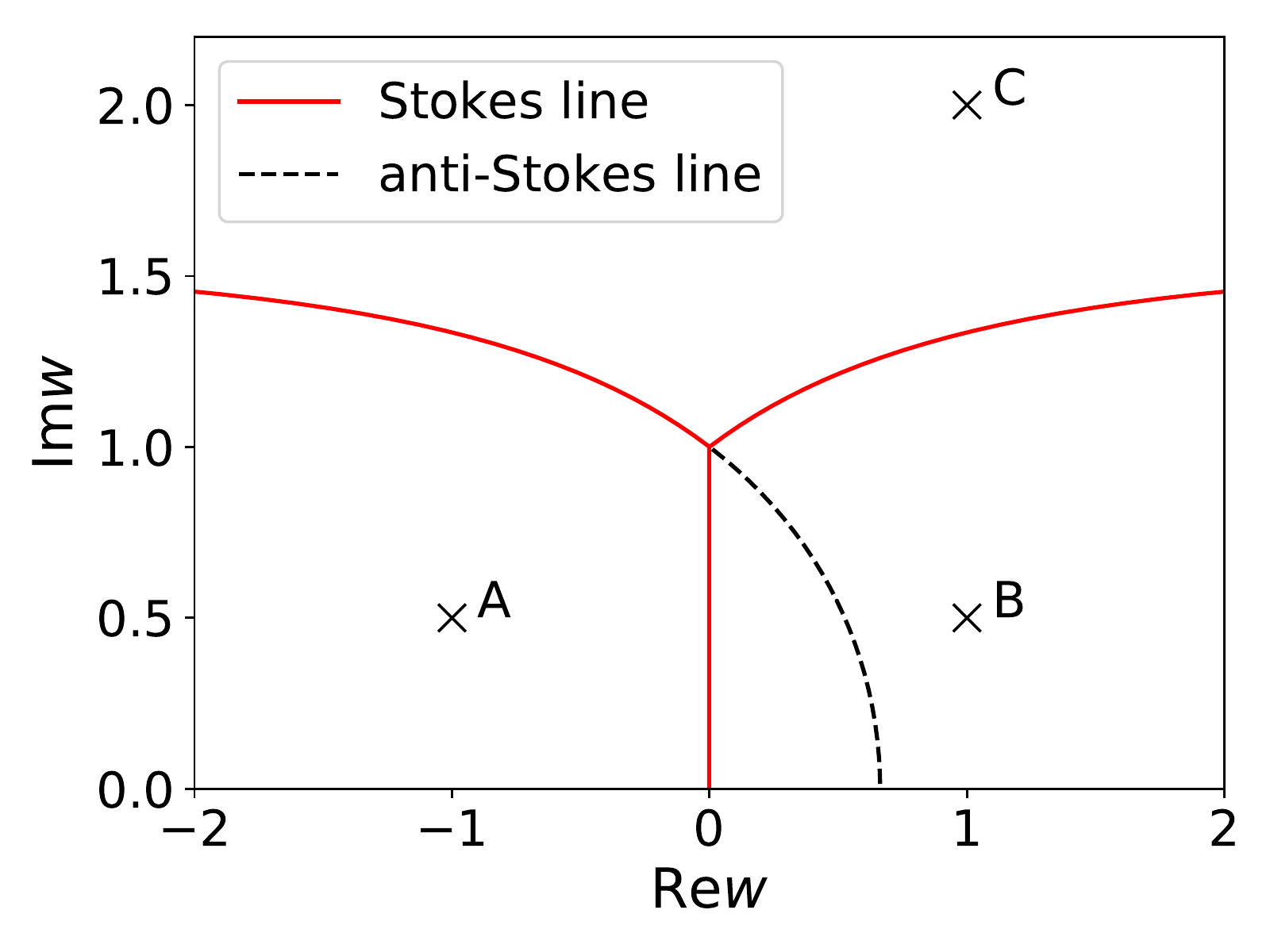}
	\fig{7.5cm}{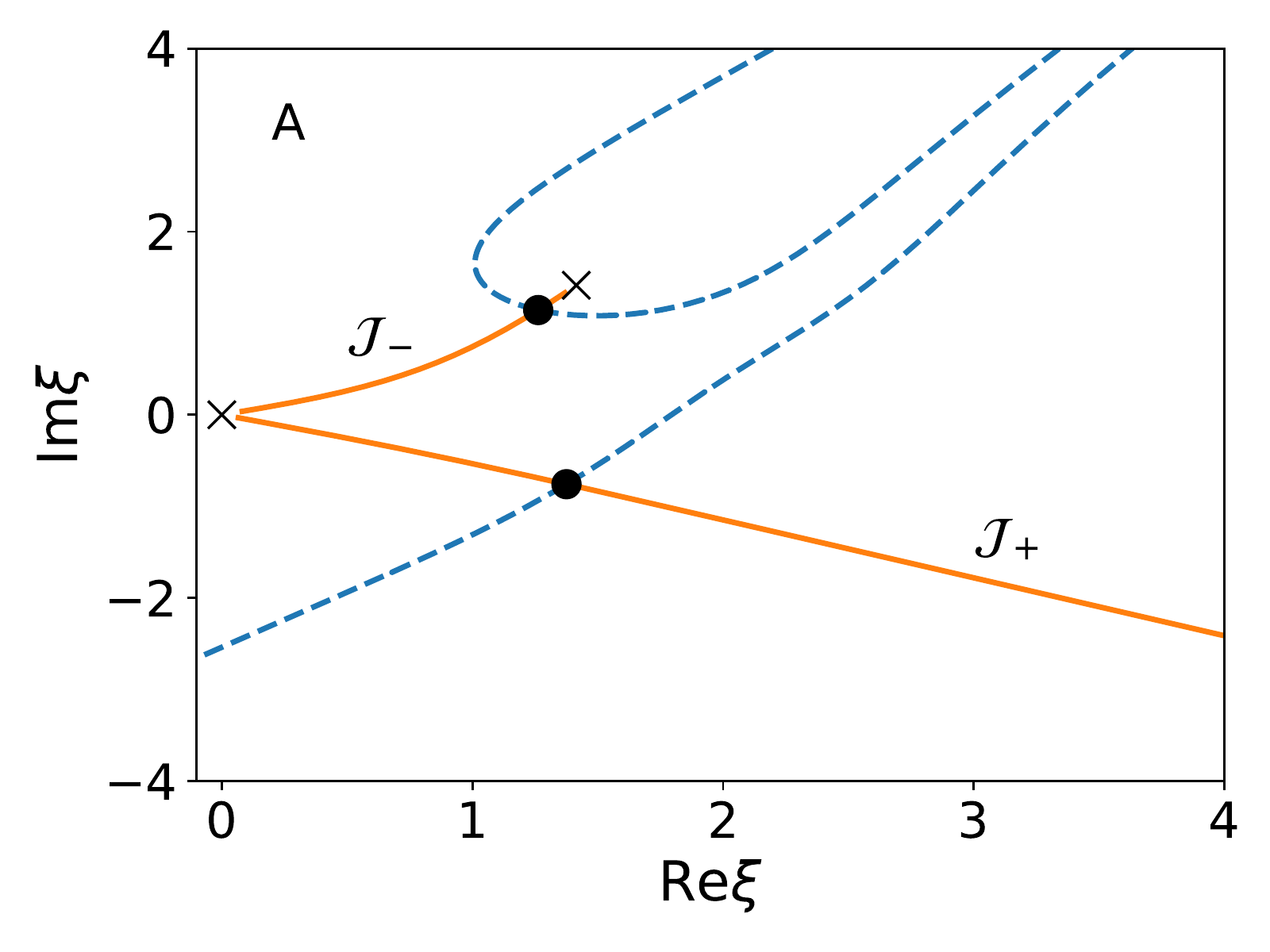}
	\fig{7.5cm}{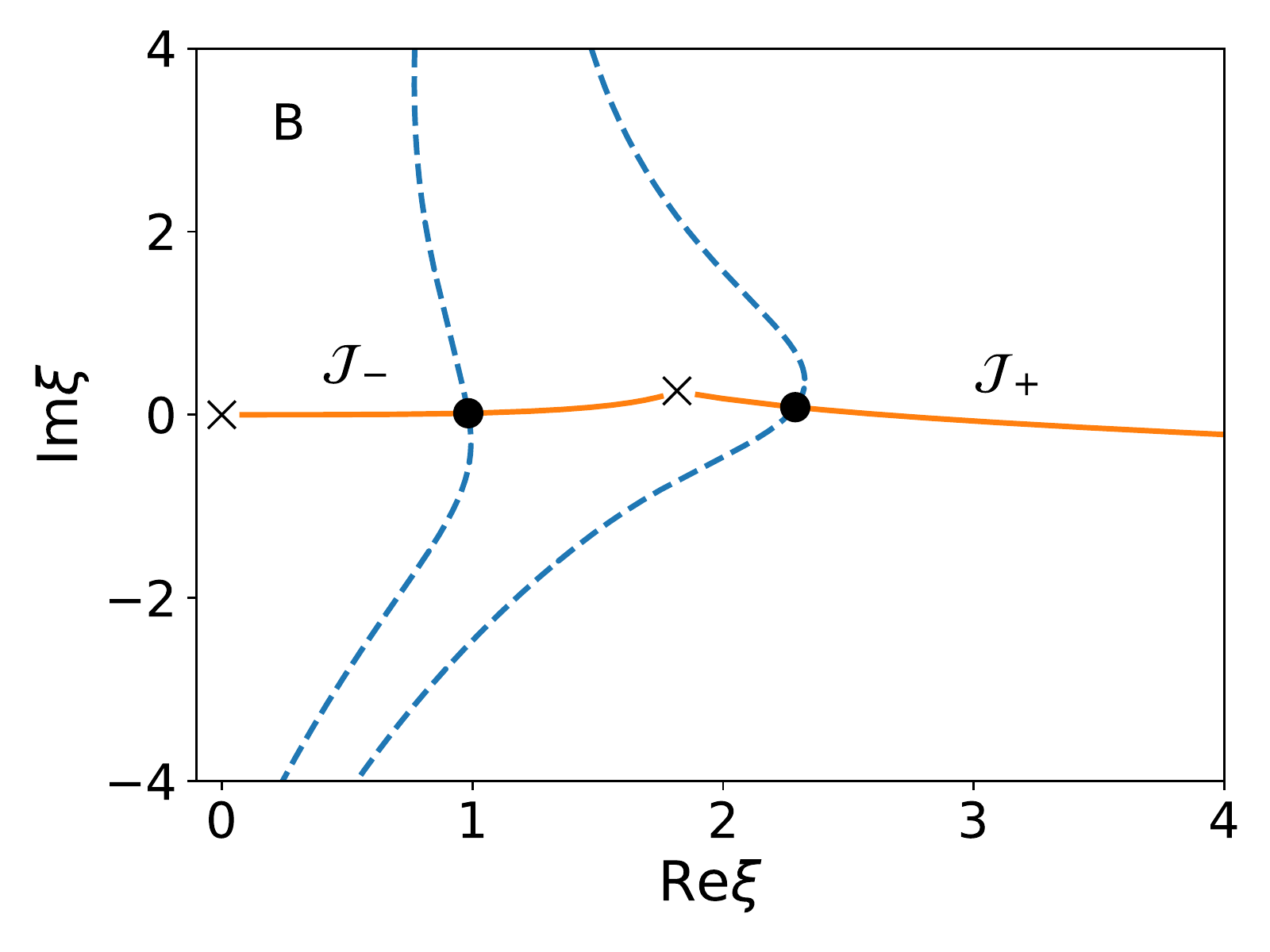}
	\fig{7.5cm}{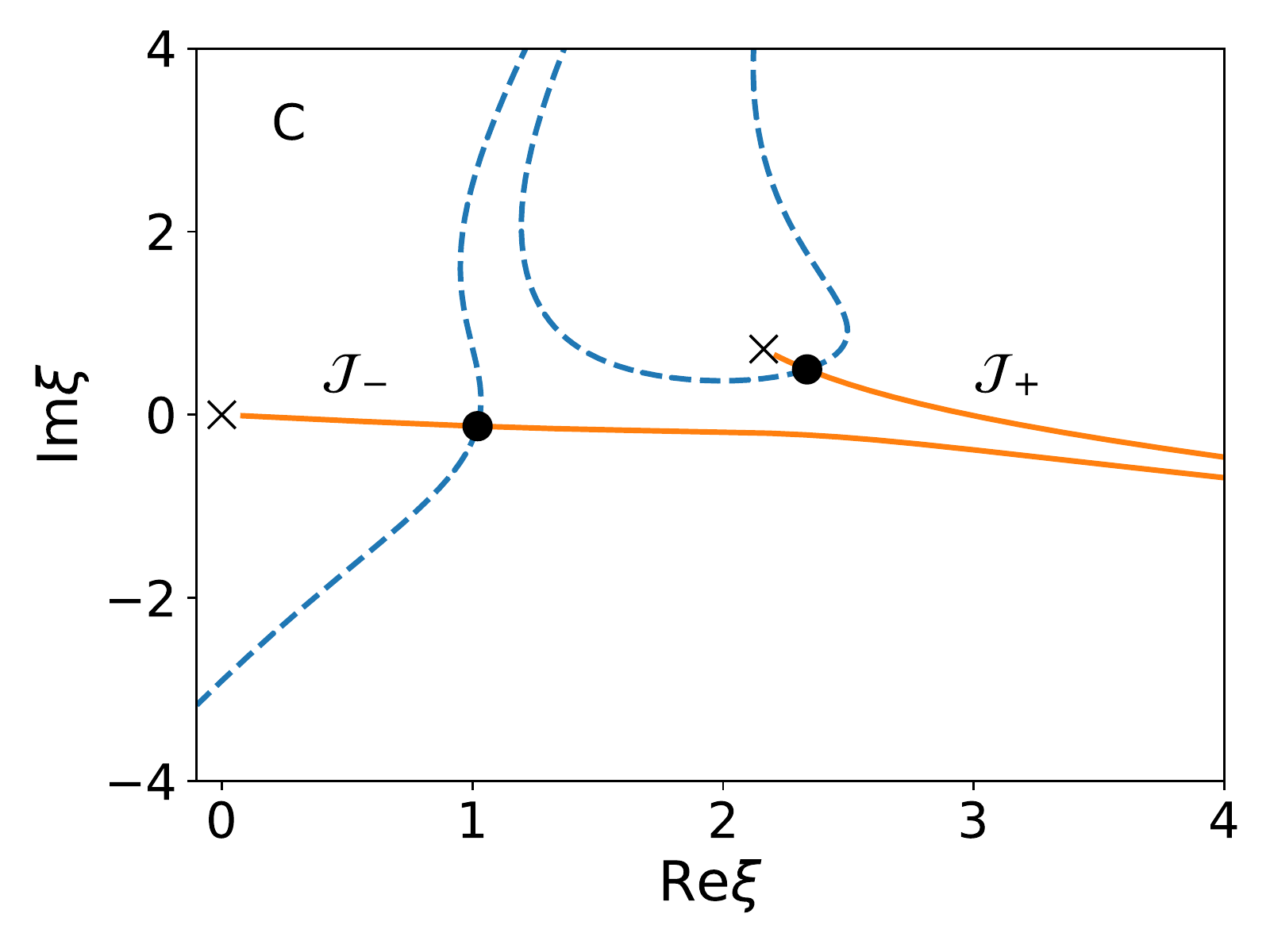}
	\caption{(Left top) The Stokes (red solid lines) and anti-Stokes lines (black dotted lines) in the complex $w$-plane.
	(Others) Lefschetz thimbles (orange solid lines) and steepest ascent contours (blue dotted lines) in the complex $\xi$-plane at $w = -1+0.5i$ (A), $1+0.5i$ (B) and $2+0.5i$ (C), respectively. Circle dots and crosses denote saddle points and singular points of the effective action $I_K$, respectively.}
	\label{fig:stokes}
\end{figure}

As we will see below,
the weight factors, which are difficult to calculate in general can be obtained explicitly in our case.
By a little algebra,
we find that the difference of the values of the effective action on the saddle points reads
\begin{align}
\Delta(w) 
\equiv 
\frac{I_K(\sigma_+) - I_K(\sigma_-)}{K}
=
2\sqrt{1 + w^2} - \log \frac{1 + \sqrt{1 + w^2}}{1 - \sqrt{1 + w^2}}.
\end{align}
Therefore, the Stokes lines are obtained as solutions of the equation\footnote{Because there are two Lefschetz thimbles in this simple example, we do not need to care about multiple of $2\pi$ ambiguity.}
\begin{align}
\Im \Delta(w) = 0. \label{stokes quartic model}
\end{align}
Since $z^{2k+1/2}a_k(z)$ is invariant under flipping the sign of the imaginary part of $z$,
one can restrict to be $\Im w \geq 0$ without loss of generality.
When $\Re w = 0$, Eq.~\eqref{stokes quartic model} has a trivial solution $w = ib$, $(0\leq b \leq 1)$.
Other solutions at $\Re w \neq 0$ can be found numerically.
We show these solutions in the upper left panel of Fig.~\ref{fig:stokes} by red solid lines.
Thus, we find that the upper half part of the complex $w$-plane is partitioned into three areas. 
From each area, we take a representative point,
say $w=-1+0.5i$ (A), $1+0.5i$ (B) and $2+0.5i$ (C),
and calculate the Lefschetz thimbles~\eqref{J} and upward flows~\eqref{K} for these values of $w$.
Other parameters, $\lambda$, $\omega^2$ and $K$ which are not relevant to this argument is fixed to $\lambda=1$, $\omega^2=1$ and $K=2$.
In the rest panels in Fig.~\ref{fig:stokes},
we show the Lefschetz thimbles by orange solid lines and upward flows by blue dotted lines.
As we discussed in Sec.~\ref{sec:thimble},
intersections of the Lefschetz thimbles and the upward flows are given by the saddle points of the effective action~\eqref{saddle}, which are denoted by circles.
Endpoints of the Lefschetz thimbles are a point at infinity or singular points of the effective action:
\begin{align}
\zeta_1=0, \ \ \ \zeta_2=\sqrt{\frac{2(z-\omega^2)}{\lambda}},
\end{align}
which are denoted by crosses.
Now, it is easy to find out the intersections of upward flows and the real axis.
As a result, we get
\begin{align}
&n_+ = 0, \quad n_- = 1, \quad
w\in \text{left bottom area (represented by A)}, \\
&n_+ = 1, \quad n_- = 1, \quad
w\in \text{right bottom area (represented by B)}, \\
&n_+ = 1, \quad n_- = 0, \quad
w\in \text{top area (represented by C)}.
\end{align}
Thanks to the above argument on the Stokes line,
the definition of the anti-Stokes line~\eqref{anti Stokes line} can be simplified as follows.
Since the Boltzmann factor $e^{-I_K(\sigma_i(w);\,w)}$ is finite,
Eq.~\eqref{anti Stokes line} has solutions if and only if both weight factors $n_-$ and $n_+$ are not zero.
In order words, $w$ should be a point in the right bottom area.
If this is the case,
Eq.~\eqref{anti Stokes line} becomes
\begin{align}
e^{-I_K(\sigma_+(w);\,w)} + e^{-I_K(\sigma_-(w);\,w)} = 0.
\end{align}
Since the relative phase between these terms is $\pi$,
we get
\begin{align}
\Re\Delta(w) = 0, \quad w\in \text{right bottom area}
\label{anti-Stokes line quartic model}
\end{align}
as a criterion for the anti-Stokes line.
This equation actually have solutions in the right bottom area,
and it forms a continuous line as depicted in the left top panel of Fig.~\ref{fig:stokes} by a black dotted line.
The endpoint of the line at $w = i$ is a point where the two saddle points are degenerated.

Before closing this subsection,
we remark that the length of the anti-Stokes line in the complex $w$-plane $L$ is bounded as
\begin{align}
 L< \frac{\pi}{2},
 \label{length of ASL}
\end{align}
and thus, $L$ is independent of the truncation order of the $\delta$-expansion $K$.
This fact seems obvious, but as we will see soon, it is crucial to argue an asymptotic behavior of the FAC condition.

\subsection{Solutions of the FAC condition and the anti-Stokes line}
\begin{figure}[tbh]
	\centering
	\fig{7.5cm}{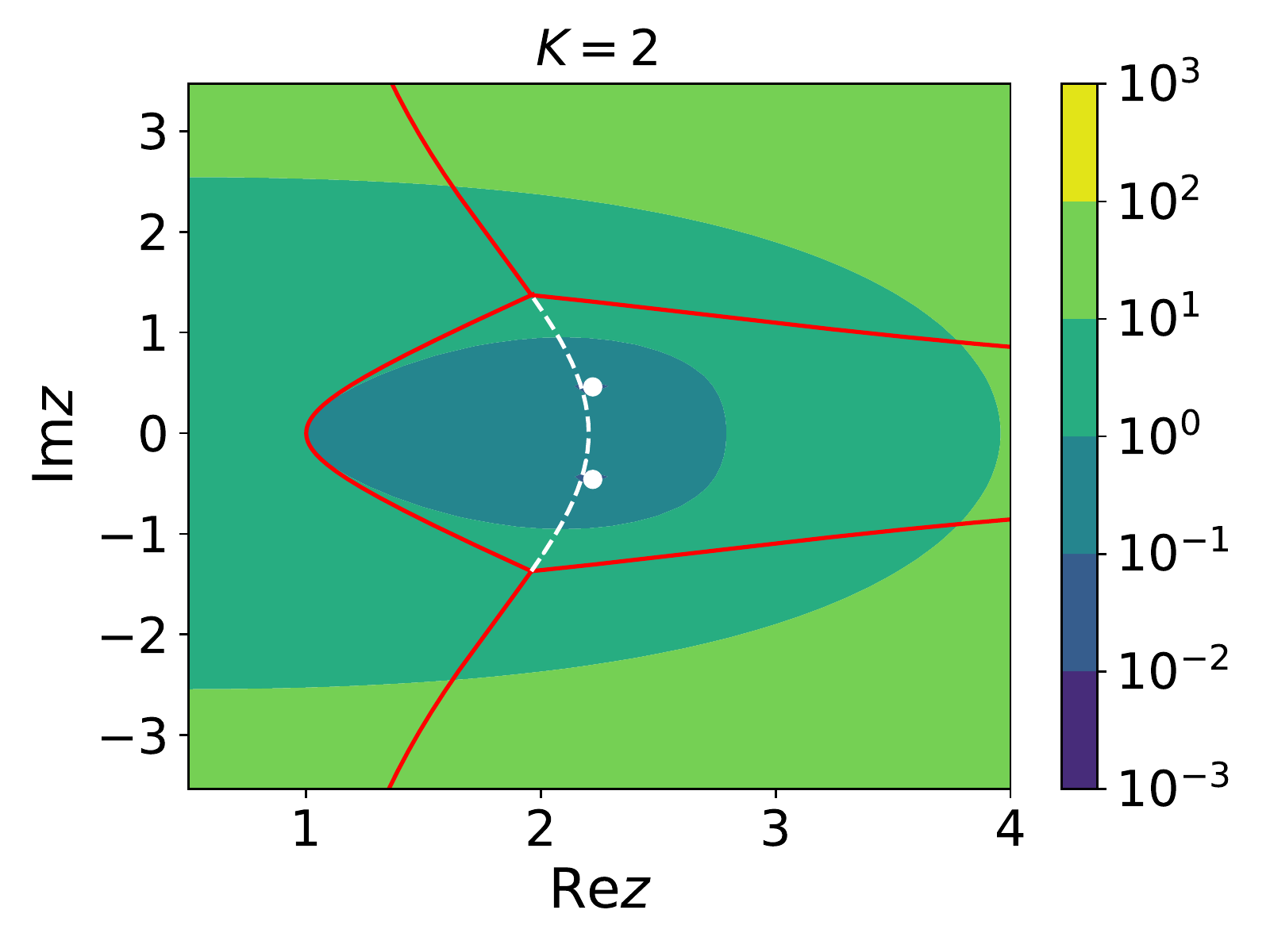}
	\fig{7.5cm}{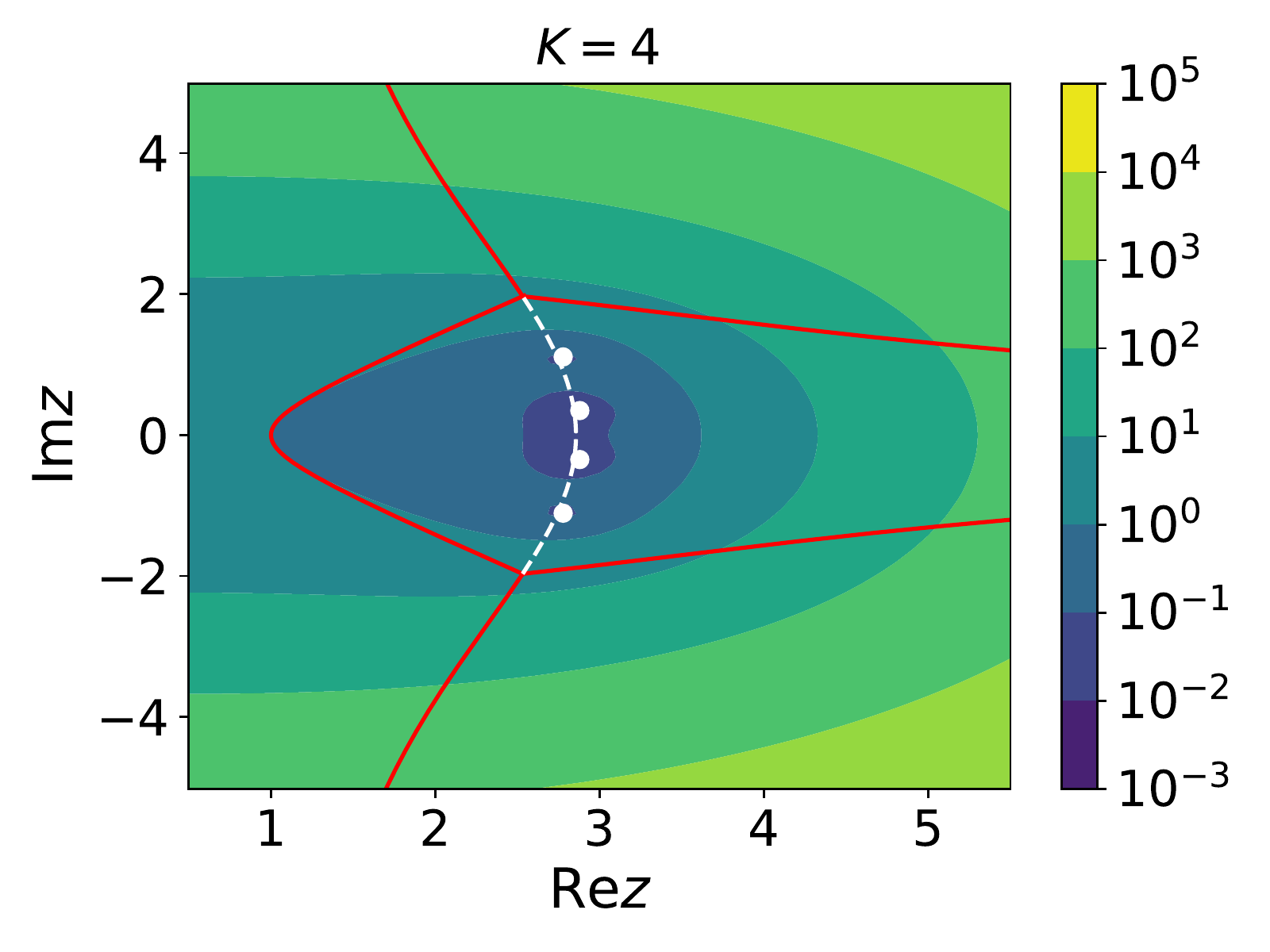}
	\fig{7.5cm}{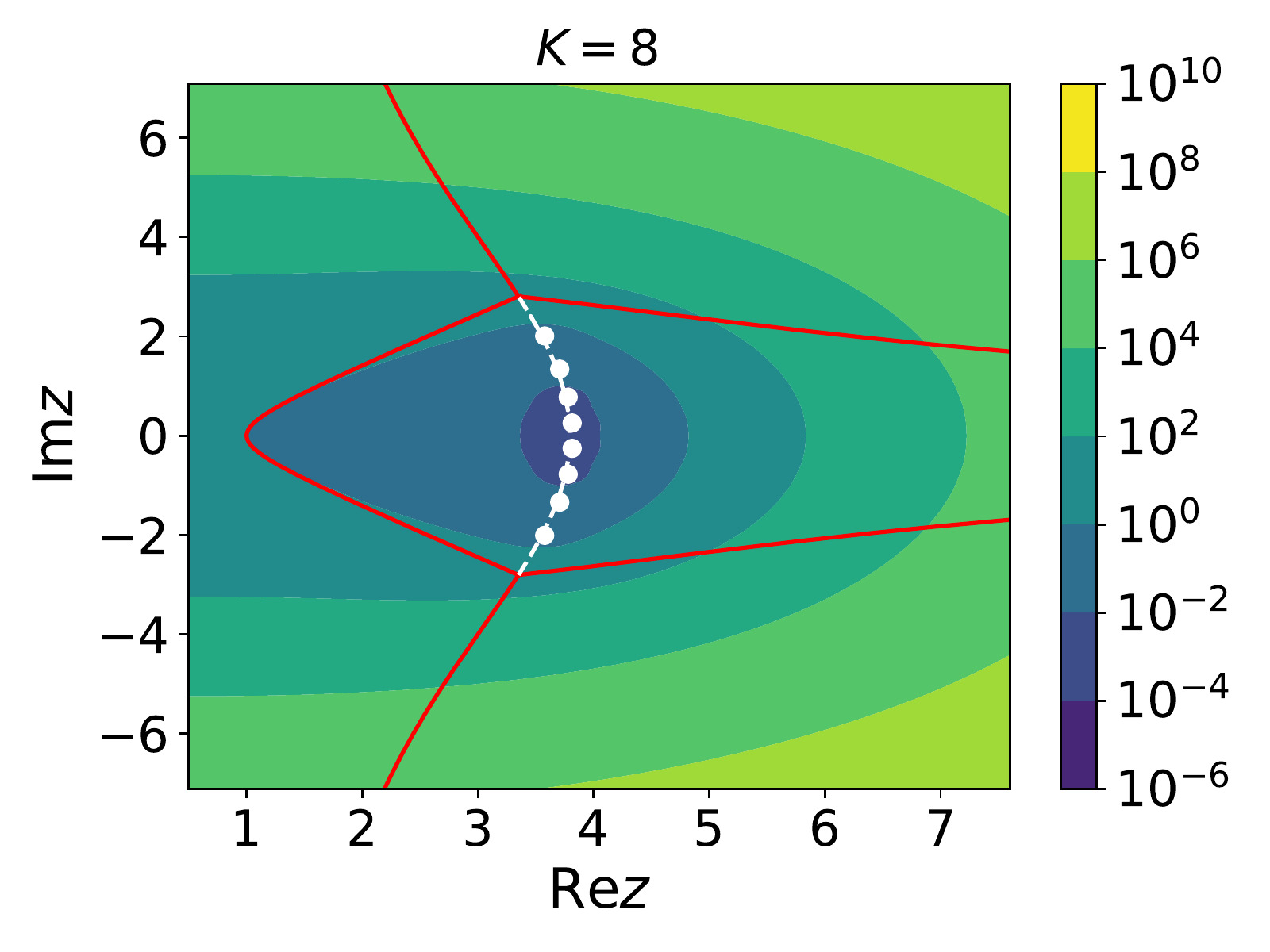}
	\fig{7.5cm}{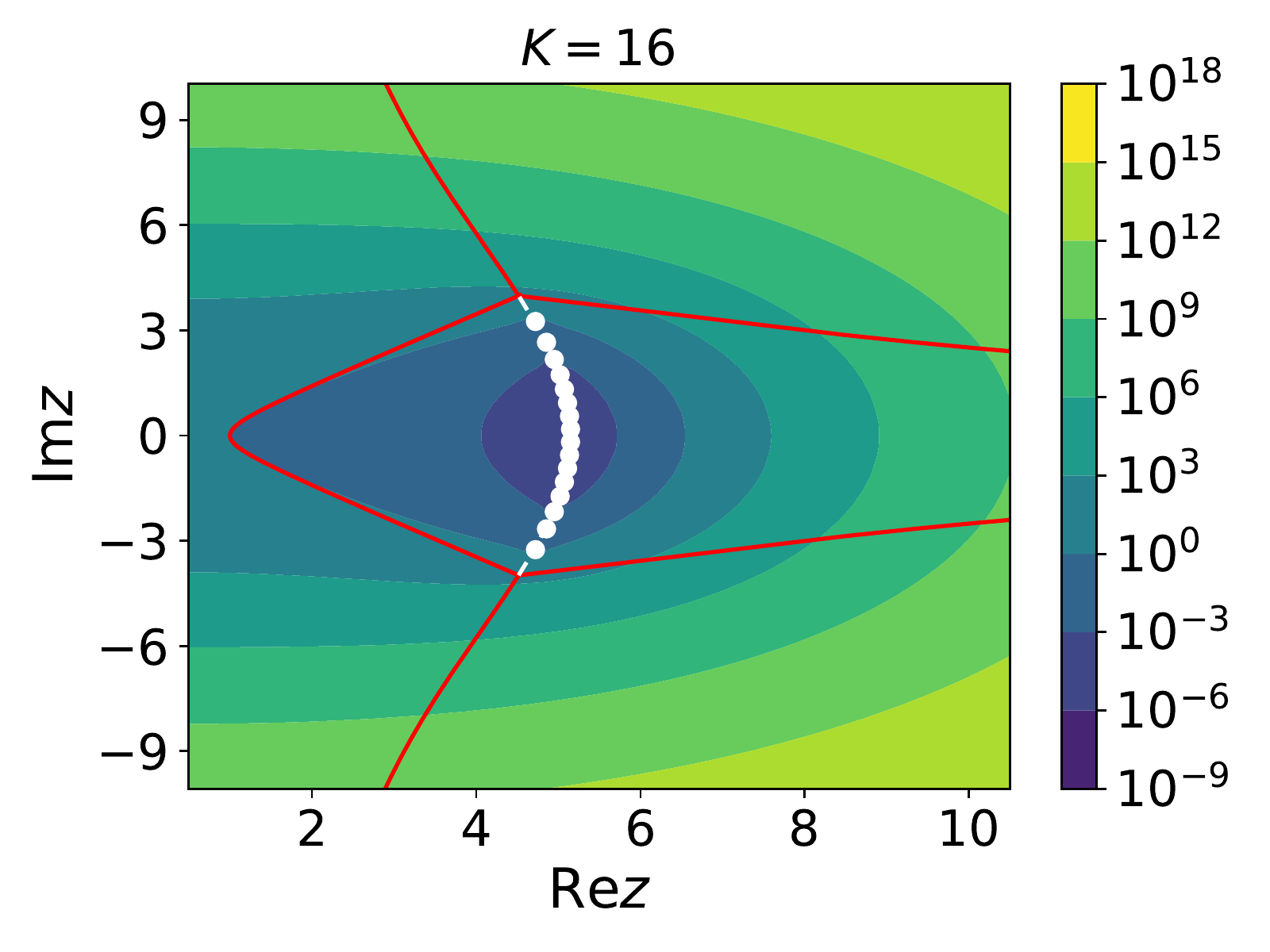}
	\caption{The solutions of the FAC conditions in the complex $z$-plane for $\lambda=1$, $\omega^2=1$ and $K=2, 4, 8, 16$ are shown by white circles.
	Red solid and white dotted lines stand for the Stokes and anti-Stokes lines, respectively. Levels of the contours is given by the absolute value of $z^{2K+1/2}a_K(z)$.}
	\label{FAC and ASL}
\end{figure}

As we discussed in Sec.~\ref{sec:thimble},
solutions of the FAC condition should be distributed around the anti-Stokes line.
This property is confirmed by Fig.~\ref{FAC and ASL},
where the solutions of the FAC condition in the complex $z$-plane are denoted by white circles for $\lambda = 1$, $\omega^2=1$ and $K=2, 4, 8, 16$.
The Stokes and anti-Stokes lines obtained in the previous subsection are mapped into the complex $z$-plane by Eq.~\eqref{z to w}, and they are denoted by red solid and white dotted lines, respectively.
We also show the absolute value of $z^{2K+1/2}a_K(z)$ by the contour plot.

A remarkable feature of $|z^{2K+1/2}a_K(z)|$ is that it tends to form a deep and flat valley around the anti-Stokes line as the truncation order of the $\delta$-expansion $K$ increases.
This tendency can be understood as follows.
Since the length of the anti-Stokes line in the complex $w$-plane is bounded as Eq.~\eqref{length of ASL},
that in the complex $z$-plane is given by $cK^{1/2}$, where $c$ is a $K$-independent constant.
On the other hand,
the FAC condition have $K$ non-degenerate solutions.
Thus, a linear density of the solutions of the FAC condition is given by
\begin{align}
\rho_K = \frac{K}{cK^{1/2}} = c^{-1} K^{1/2}.
\end{align}
This means that the solutions of the FAC condition accumulate on the anti-Stokes line in the limit $K\to\infty$.
In this limit, the FAC condition holds everywhere as long as $z\propto K^{1/2}$ due to the identity theorem~\footnote{
The scaling behaviour $z\propto K^{1/2}$ agrees with the previous proofs of the convergence of the $\delta$-expansion.}.
This result is natural 
because the approximant of the $\delta$-expansion $Z_K$ should be insensitive to the choice of the unphysical parameter $z$. 

Similar discussion can be performed for a general $q$, which is an exponent of the highest order term of the general effective action Eq.~\eqref{effective action: factorized form}.
In this case, the $K$-dependence of $z$ should be $z \propto K^{1-2/q}$ so that the saddle point approximation is justified.
Once $K$ is factorized from the effective action, 
the Stokes and anti-Stokes lines should be determined irrespective of $K$.
Thus, the density of the solutions of the FAC condition around the anti-Stokes line would be given by
\begin{align}
\rho_K = \frac{K}{cK^{1-2/q}} = c^{-1} K^{2/q}.
\end{align}
Therefore, as long as $q$ is finite,
the FAC condition holds everywhere due to the accumulation of the solutions.
The only exception is that one considers infinitely large $q$.
In this case, physical quantities could be sensitive to a choice of the parameter $z$. 

\begin{figure}[tbh]
	\centering
	\fig{7.5cm}{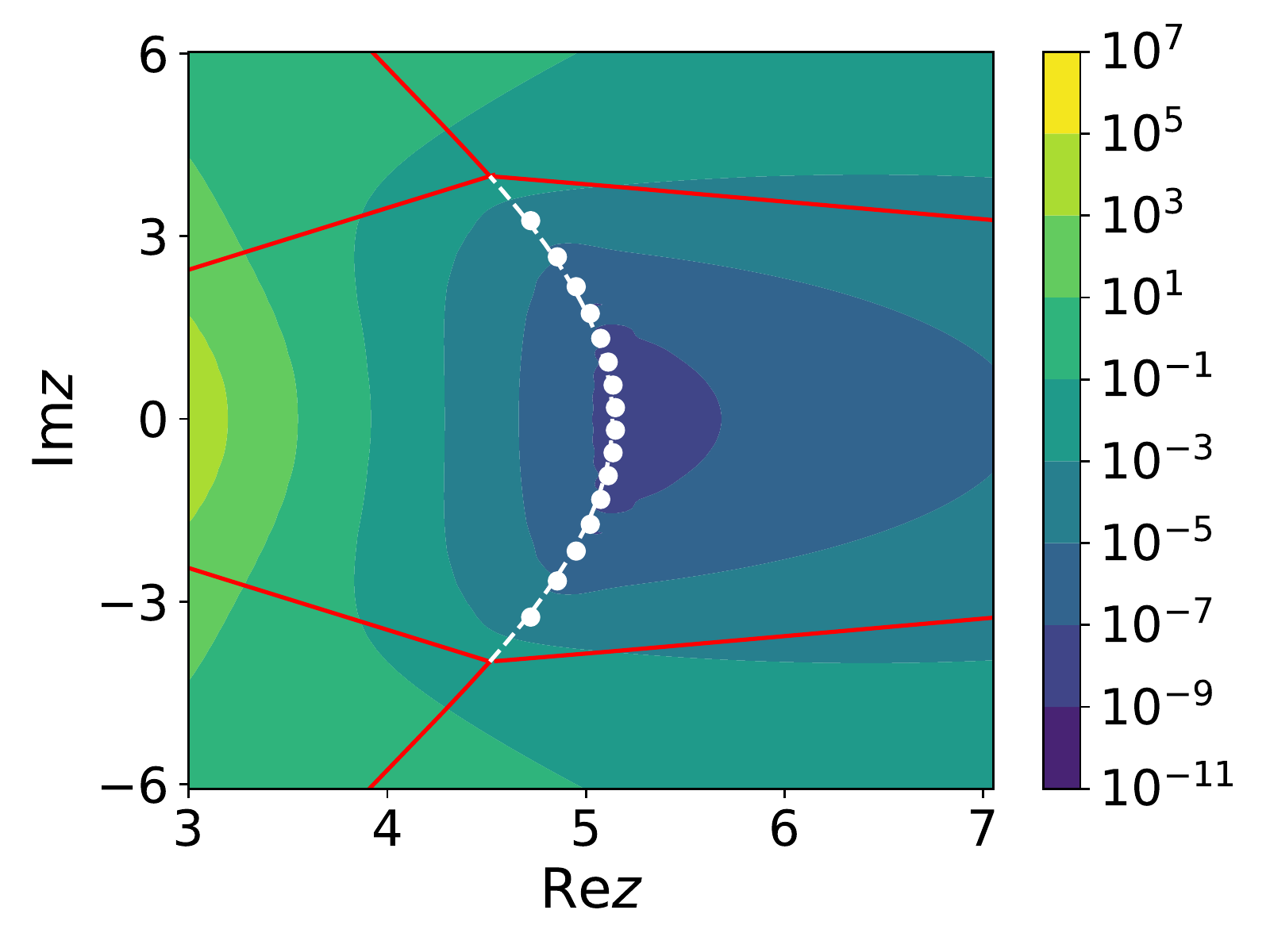}
	\fig{7.5cm}{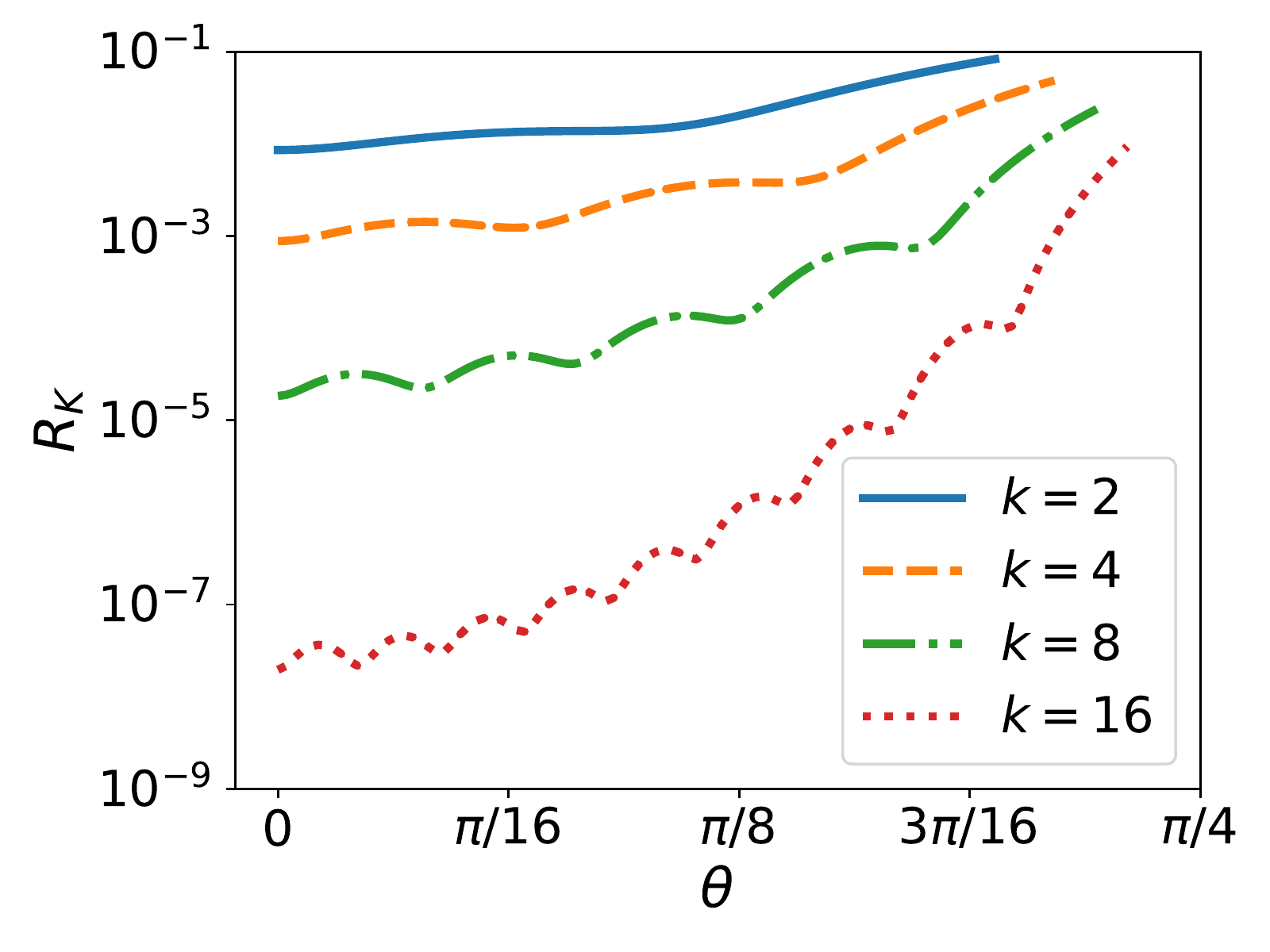}
	\caption{
	The value of $R_K$ at $K=16$ in the complex $z$-plane (left) and at $K=2,4,8$ and 16 on the anti-Stokes line (right).
	In the right panel, $\theta$ denotes the angle of the anti-Stokes line.
	In both panels, $\lambda=1$, $\omega^2=1$. 
	Symbols and lines are same as Fig.~\ref{fig:stokes}.
    }
	\label{fig:RK}
\end{figure}

In the rest part of this subsection,
we discuss how physical quantities depends on the parameter $z$ around the anti-Stokes line.
To see this, 
we depict the reminder of the $\delta$-expansion
\begin{align}
R_K(z) \equiv |Z-Z_K(z)|,
\label{reminder}
\end{align}
in the left panel of Fig.~\ref{fig:RK} for $K=16$.
We also put $\lambda=1$ and $\omega^2=1$, which are same values as used for Fig.~\ref{fig:stokes}.
Again, we find a deep and flat valley around the anti-Stokes line similar to that of $|z^{2K+1/2}a_K(z)|$.
In order to visualize more precise structure of the valley and its $K$ dependence, 
we plot $R_K$ along the anti-Stokes line for $K=2,4,8$ and 16 in the right panel of Fig.~\ref{fig:RK}
In this figure, the horizontal axis shows the angle of the anti-Stokes line.
Even when $z$ is complex, the order of magnitude of $R_K$ decreases as $K$ becomes large.
We obtain similar results for different $\lambda$ and $\omega^2$ as shown in appendix~\ref{App:RK}.

In Fig.~\ref{fig:plateau}, 
we compare $K$th-order approximants $Z_K$ along the real axis and $\arg z = \pi/8$ line with the exact value.
The horizontal axis is shifted as $z - z_\text{asl}$, where $z_\text{asl}$ is the point of the anti-Stokes line.
In both figures,
$Z_K$ forms a plateau around $z_\text{asl}$ as $K$ increases, and thus, it becomes insensitive to the choice of the unphysical parameter $z$.
In many applications of OPT, this kind of plateau structure is used as a guiding principle to determine an optimal value of $z$.
What we clarified here is that the underlying nature of the plateau formation is an accumulation of the solutions of the FAC conditions in the complex $z$-plane.

\begin{figure}[tbh]
	\centering
	\fig{7.5cm}{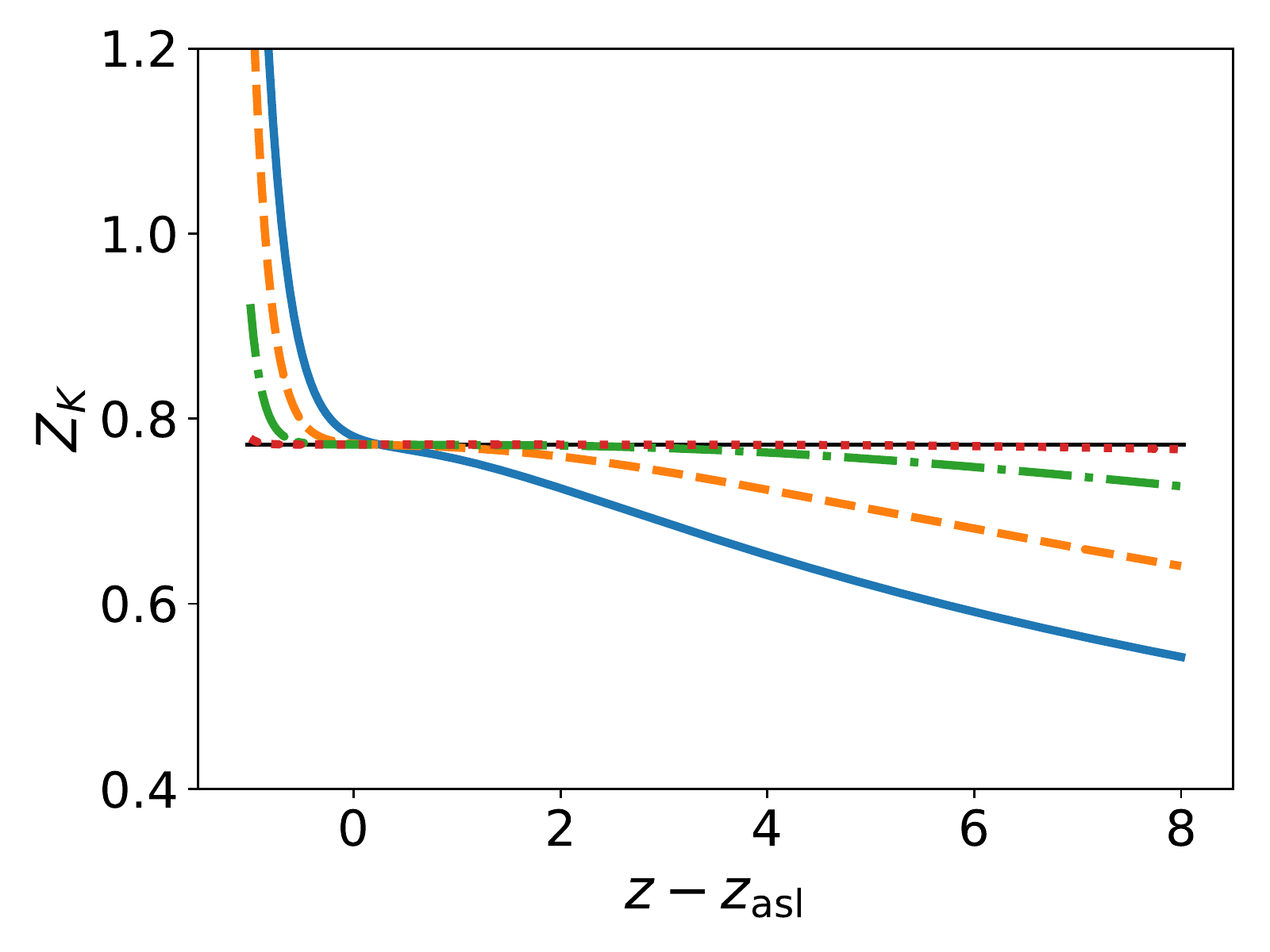}
	\fig{7.5cm}{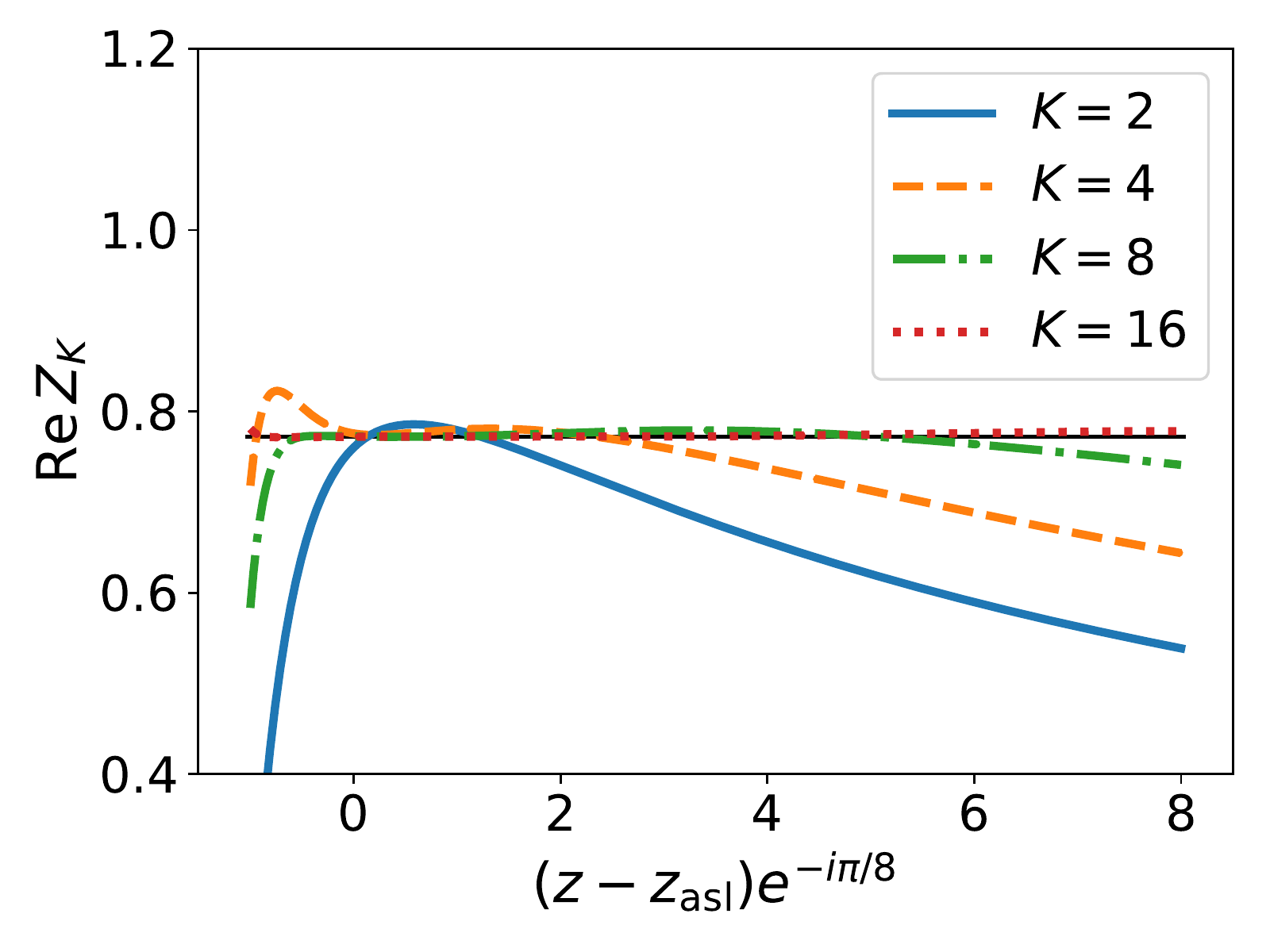}
	\caption{
    The $z$-dependence of the $K$th-order approximants $Z_K$ on the real axis (left) and $\arg z=\pi/8$-line around the anti-Stokes line. 
    The horizontal line shows the exact value of $Z$.
    }
	\label{fig:plateau}
\end{figure}

\subsection{Relation to the PMS condition}
In the previous subsection,
we see that the plateau structure of the approximant of the physical quantity $Z_K$ emerges around the anti-Stokes line determined by the FAC condition.
However, strictly speaking, this is a quite nontrivial phenomenon because the FAC condition is a condition for the highest order term of the $\delta$-expansion, not for the physical quantity itself.
Here, we show why this phenomenon happens through the relation to the PMS condition, which could detect the plateau structure of $Z_K$ more directly.

As pointed in~\cite{Buckley:1992pc},
the first-order derivative of $Z_K$ with respect to $z$ also has a simple integral form:
\begin{align}
\frac{\partial Z_K(z)}{\partial z}
&=
\frac{(-1)^{K+1}}{2K!}
\int_{-\infty}^\infty\frac{dx}{\sqrt{2\pi}}
\left(\frac{\omega^2-z}{2}x^2 + \frac{\lambda}{4}x^4 \right)^K x^{2}
e^{-z x^2/2} \\
&=
\frac{(-1)^{K+1}}{2K!}
\int_{-\infty}^\infty\frac{dx}{\sqrt{2\pi}}
e^{-I_K^\text{PMS}(x;z)},
\label{PMS: integral rep.}
\end{align} 
where an effective action for the PMS condition $I_K^\text{PMS}(x;z)$ reads
\begin{align}
I_K^\text{PMS}(x;z) 
=
\frac{zx^2}{2} 
- K\log \left(\frac{\omega^2-z}{2}x^2 + \frac{\lambda}{4}x^4 \right)
- \log x^2.
\label{effective action: PMS}
\end{align}
Equation~\eqref{effective action: PMS} involves an additional term $\log x^2$ compared with the effective action for the FAC condition~\eqref{effective action: FAC}.
This term only gives a subleading contribution in the saddle point approximation after the rescaling $z \to K^{1/2}z$ and $x \to K^{1/4}x$ discussed in Sec.~\ref{sec:thimble}.
Therefore, the distribution of solutions of the PMS condition coincide with the anti-Stokes line determined by the FAC condition in the limit $K \to \infty$.

In addition, 
one can estimate locations of the solutions of the PMS condition at finite $K$ based on a geometrical argument.
Comparing Eq.~\eqref{PMS: integral rep.} with Eq.~\eqref{ak integral expression}, we find
\begin{align}
\frac{\del a_{K+1}}{\del \omega^2}
=
\frac{\del Z_K}{\del z}.
\end{align}
By using the analytic expression of $a_K$~\eqref{ak analytic expression},
the PMS condition can be rewritten as
\begin{align}
\frac{\del}{\del\omega^2}
{}_1F_1\lp-K, \frac{1}{2}-2K; \frac{(\omega^2-z)z}{\lambda}\rp
=0,
\end{align}
or equivalently
\begin{align}
\frac{z}{\omega^2-2z} 
\frac{\del}{\del z}
{}_1F_1\lp-K, \frac{1}{2}-2K; \frac{(\omega^2-z)z}{\lambda}\rp
=0.
\end{align}
Thus, we find that the solutions of the PMS condition are given by those of the first-order derivative of the FAC condition.
Thanks to this relation,
we can constrain the distribution of the solutions of the PMS condition by the following argument.
Let us put $\zeta = (\omega^2-z)z/\lambda$, 
and calculate zeros of $P(\zeta) \equiv {}_1F_1\lp-K, \frac{1}{2}-2K; \zeta\rp$.
Because $P(\zeta)$ is a polynomial function of $\zeta$,
all zeros of $P'(\zeta)$ lie in the convex hull of the set of the zeros of $P(\zeta)$~\footnote{This is known as Lucas's theorem.}.
These are depicted in the left panel of Fig.~\ref{PMS}.
Mapping the boundary of the convex hull on the $\zeta$-plane to the $z$-plane,
we obtain the allowed region where the solutions of the PMS condition can appear as shown in the right panel of Fig.~\ref{PMS}.
Since the location of the boundary of the convex hull has the same $K$-dependence as that of the anti-Stokes line\footnote{Specifically, the dependence is gevin by $K^{1/2}$.},
this constraint ensures that the solutions of the PMS condition also distribute around the anti-Stokes line determined by the integral representation of the FAC condition. 
Therefore, we conclude that physical quantities are insensitive to $z$ as long as $z$ is chosen to be close to the anti-Stokes line.
\begin{figure}[tbh]
	\centering
	\fig{7.5cm}{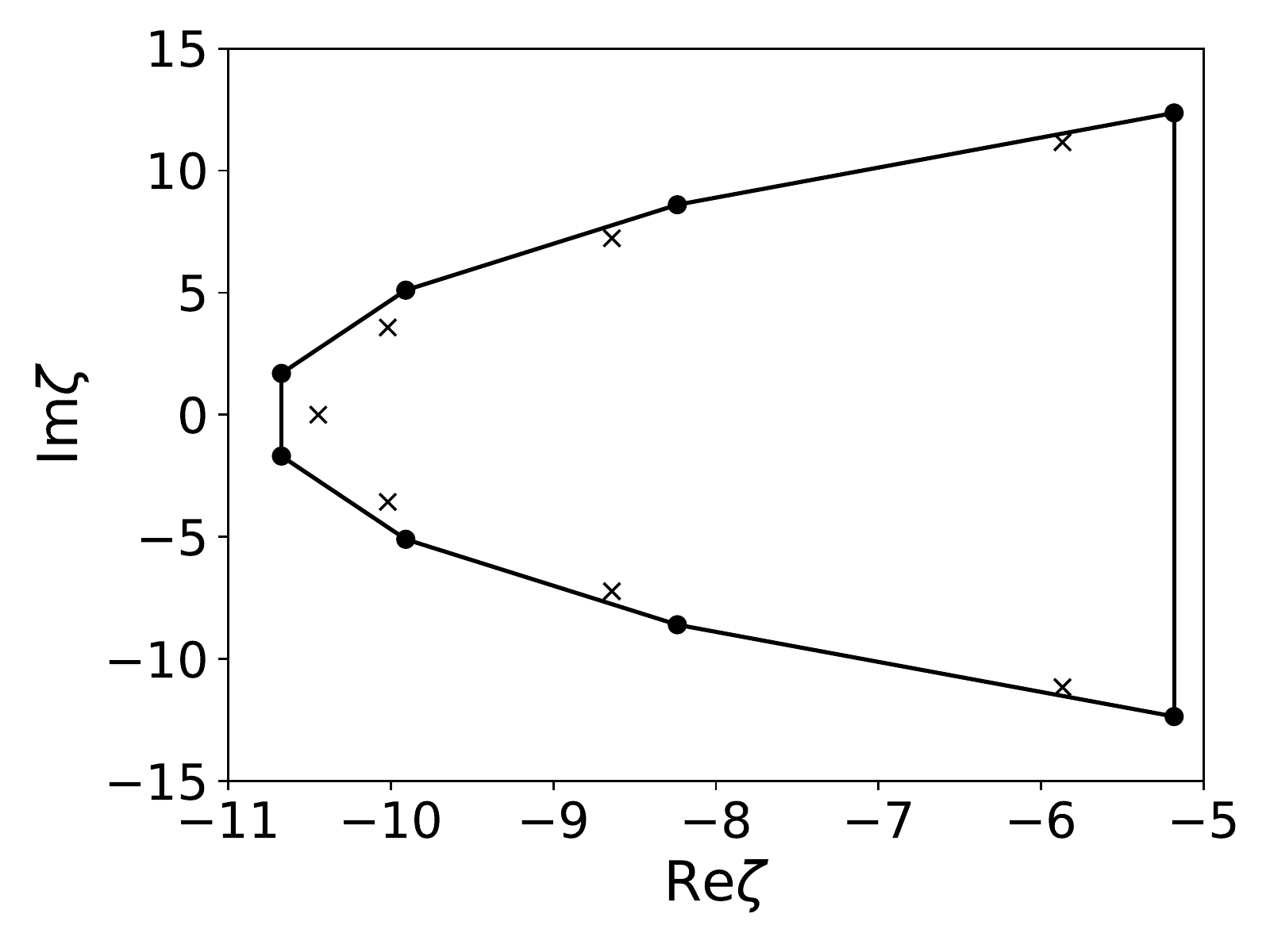}
	\fig{7.5cm}{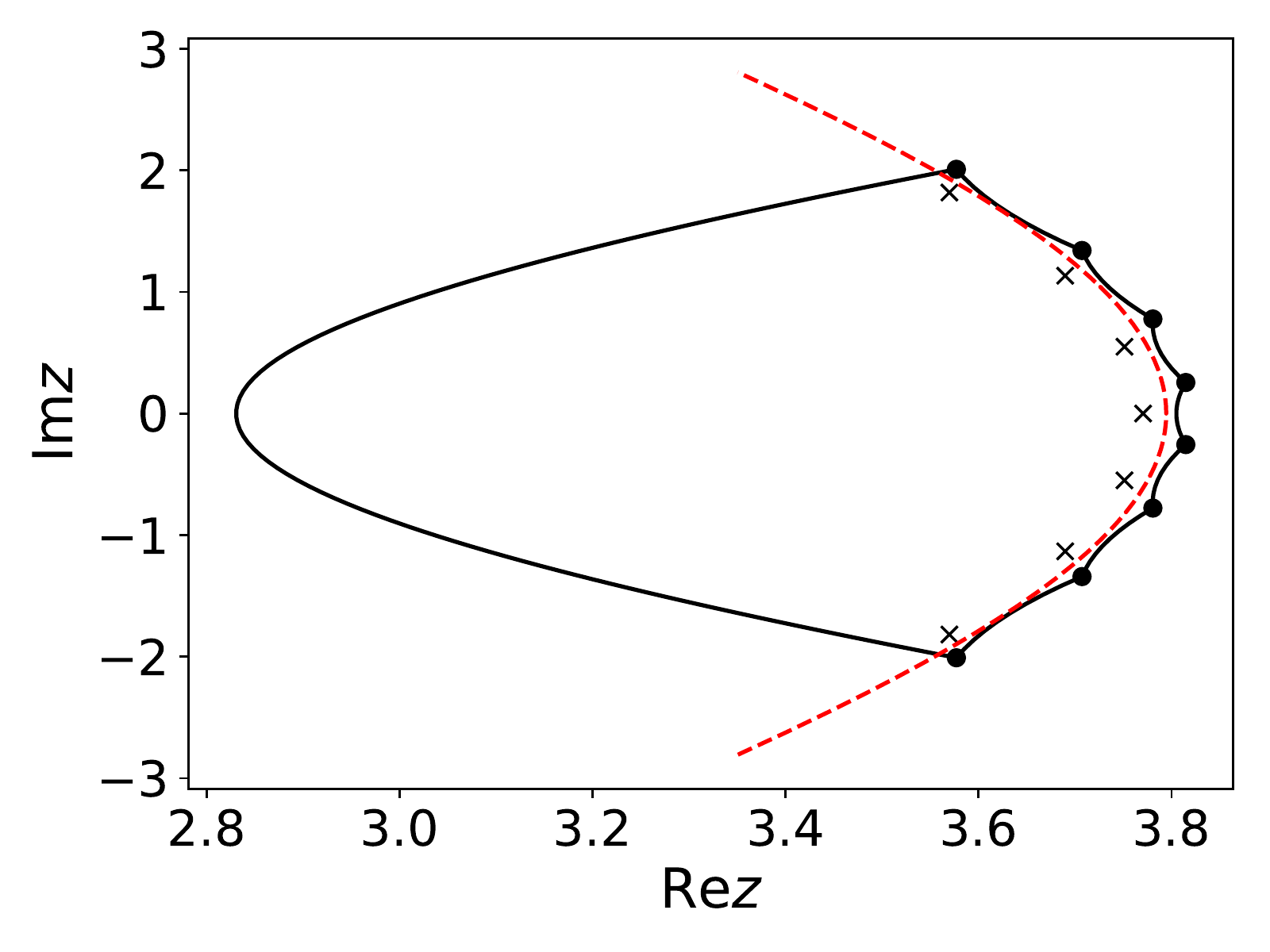}
	\caption{(Left) Zeros of $P$ and $P'$ are shown by circles and crosses for $K=8$.
    The solid lines shows the convex hull of the set of the zeros of $P$.
    (Right) The left figure obtained in the $\zeta$-plane is converted to the $z$-plane for $\lambda=1$ and $\omega^2=-1$.
    Th red dotted line shows the anti-Stokes line.
    }
	\label{fig:PMS}
\end{figure}

Our argument here would be useful to practical applications of OPT, where one cannot obtain higher order terms of perturbation theory.
As demonstrated in Fig.~\ref{fig:plateau},
the plateau structure is not clear when $K$ is small.
In that case, the PMS condition is useless to determine $z$.
Nevertheless, 
the FAC condition could give a rigorous criterion for $z$.

\subsection{Analogy to the statistical physics}
In this subsection, 
we point out that there is a clear analogy between 
the accumulation of the solutions of the FAC condition
and phase transitions in the statistical physics.

According to a renown work by Lee and Yang, 
zeros of partition function in a complex parameter plane, which are referred to as Lee-Yang zeros have rich information about phase transitions~\cite{Yang:1952be, Lee:1952ig}. 
In general, 
a set of the Lee-Yang zeros forms a continuous arc in the complex parameter plane in the thermodynamics limit.
If the arc pinches the real axis, 
a free energy has a singularity, and this means a first order phase transition.
Thus, 
the distribution of the Lee-Yang zeros, namely that in the thermodynamic limit is crucial for arguments on the phase transition.
It is known that the distribution can be also well described by anti-Stokes line.
These properties are well studied in the Ising model and the mean field Gross-Neveu model and so on.
Based on the picture,
the pinching is understood as accumulation of the Lee-Yang zeros on the anti-Stokes line~\cite{Itzykson:1983gb, Pisani1993, Kanazawa:2014qma}.

This argument is parallel to that in the previous sections.
In the case of OPT, 
the zeros of the $K$th-order term $a_K(z)$ of the $\delta$-expansion and the solutions of the FAC condition corresponds to the partition function and Lee-Yang zeros, respectively.
Thus, 
the procedure to find an optimal $z$ based on the FAC condition can be regarded as the problem to find the first order transition in the complex $z$-plane. 
The analogy between OPT and the statistical physics 
are summarized in Table~\ref{analogy}. 

On the other hand,
there is a difference between OPT and the statistical physics.
While only the zeros on the real axis have physical meaning in arguments of the first order phase transition, 
the solution of the FAC condition is not necessarily real even when the coupling constant is real.
Moreover, the best choice of $z$ in no longer given by the intersection of the anti-Stokes line and the real axis when the coupling constant is complex.
These are demonstrated in App.~\ref{App:RK}.
\begin{table}[th]
	\caption{An analogy.}
	\label{analogy}
	\centering
	\begin{tabular}{cc}
		\hline
		Optimized perturbation theory & Statistical physics\\
		\hline
		$K$th-order coefficient $a_K$ & Partition function\\
		Optimal $z$ & Lee-Yang zero\\
		$K \to \infty$ limit & Themodynamic limit \\
		\hline
	\end{tabular}
\end{table}

\section{Summary and Concluding remarks}\label{sec:summary}
In this paper, 
we have studied fundamental properties of the FAC condition which is used as a variational criterion in OPT.
We have pointed out that the FAC condition has two conceptual problems.
One is that we need to put additional criteria by hand to determine the artificial parameter since a solution of the FAC condition is not unique.
The another is that the insensitivity of the approximant as a function of an artificial parameter which is not guaranteed.

We have clarified that a distribution of the solutions of the FAC condition is related to topology of Lefschetz thimbles.
As a result, we have found that the solutions are distributed around the anti-Stokes line on which contributions from each thimble cancel out with each other relying on the saddle point approximation.
Moreover, we have argued that the approximation becomes exact in the limit $K\to \infty$, where $K$ is a truncation order of the reorganized perturbative expansion.

We have performed detailed studies on the anti-Stokes line for a one dimensional integral.
In that concrete example,
the saddle point approximation makes sense when we assume that the artificial parameter $z$ is scaled as $z \propto K^{1/2}$.
Then, we have shown that the length of the anti-Stokes line in the complex $z$-plane is also proportional to $K^{1/2}$.
Since the FAC condition has $K$ non-degenerated solutions, we have concluded that these solutions accumulate on the anti-Stokes line in the $K\to\infty$ limit.
If this is the case, the FAC condition is satisfied for any $z$ as long as $z \propto K^{1/2}$.
This is an underlying mechanism that physical quantities calculated by OPT can be insensitive to the choice of $z$. 
We have confirmed that an approximated quantity obtained in OPT agrees with the exact value along the anti-Stokes line and a flat region is developed around the line as $K$ increases.

In addition, we have also discussed that the relation between the FAC and PMS conditions in order to clarify why the FAC condition leads the insensitivity of the physical quantities.
We have found that the PMS condition also has a same effective action as that for the FAC condition in the limit $K\to\infty$.
In addition, we have argued that the solutions of the PMS conditions must be interior points of the convex hull of the solutions of the FAC condition, 
and this ensures that the solutions of the PMS conditions always appear around the anti-Stokes line determined by the FAC condition.

Finally, we have pointed out that there is a clear analogy between our argument and the physics of phase transitions.
According to the analogy, a partition function and the thermodynamic limit corresponds to the integral representation of the FAC condition and $K\to\infty$ limit, respectively.

While we have restricted our attention to one dimensional integrals in this paper,
the arguments performed here can be generalized to higher-dimensional integral, and presumably to path integrals.
Actually, calculating topology of Lefschetz thimbles in infinite dimensions is a formidable task.
However, we expect that it turns to be tractable once we reformulate OPT as the physics of phase transitions in terms of the artificial parameter $z$, where many familiar tools and techniques are available even in quantum field theories.
We will present these discussion elsewhere.

\section*{Acknowledgments}
T.M.D. is supported by 
the RIKEN Special Postdoctoral Researchers Program.

\appendix

\section{Error estimation of the saddle point approximation}\label{App:saddle point}
In this appendix, 
we prove that the FAC condition is given by Eq.~\eqref{anti Stokes line} in the limit $K\to\infty$.
The factorized form of the effective action~\eqref{effective action: factorized form} is given by
\begin{align}
	I_K(\xi)
	&=
	K \left[
	\frac{z}{2}\xi^2 
	- \log \lp -\frac{z}{2}\xi^2 + \lambda_q \xi^q + O(K^{-1/q}) \rp
	\right] \\
	&=
	K \left[
	\frac{z}{2}\xi^2 
	- \log \lp -\frac{z}{2}\xi^2 + \lambda_q \xi^q \rp + r_K
	\right] \equiv K \tilde{I}(\xi),
\end{align} 
where $r_K$ has a form of
\begin{align}
r_K =  \lp -\frac{z}{2}\xi^2 + \lambda_q \xi^q \rp \times O(K^{-1/q}).
\end{align}
Here, we omit the constant term.
A saddle point of the effective action is given by
\begin{align}
z\xi - \frac{-z\xi + q\lambda_q\xi^{q-1}}{-\frac{z}{2}\xi^2 + \lambda_q \xi^q} + \frac{O(K^{-1/q})}{\lp-\frac{z}{2}\xi^2 + \lambda_q \xi^q\rp^2} = 0.
\end{align}
Since the saddle point is away from the zeros of $-\frac{z}{2}\xi^2 + \lambda_q \xi^q$, 
the last term of the right hand side is actually suppressed as $K^{-1/q}$ compared with the other terms.
Thus, the saddle point also has a form:
\begin{align}
	\sigma = \sigma_0 + O(K^{-1/q}).
	\label{saddle point at leading order}
\end{align}
By expanding the effective action around the saddle point,
an integral on each Lefschetz thimble reads~\cite{Fedoryuk1977, Fedoryuk1989}
\begin{align}
	\int_{\calJ_i} d\xi e^{-I_K(\xi)}
	&=
	\sqrt{\frac{2\pi}{K}}
	\exp\left[-K\tilde{I}(\sigma) -\frac{1}{2}\tilde{I}''(\sigma) \lp 1 + O(K^{-1})\rp \right].
\end{align}
Due to Eq.~\eqref{saddle point at leading order},
the second term involved in the exponential function becomes
\begin{align}
\tilde{I}''(\sigma)
=
\tilde{I}''(\sigma_0) + O(K^{-1/q}),
\end{align}
and it is at most $O(1)$.
Therefore, each integral on a thimble is given by $\sqrt{\frac{2\pi}{K}}e^{-K\tilde{I}(\sigma)}$ in the limit $K\to\infty$.

\section{$R_K$ for other parameters}\label{App:RK}
\begin{figure}[tbh]
	\centering
	\fig{7.5cm}{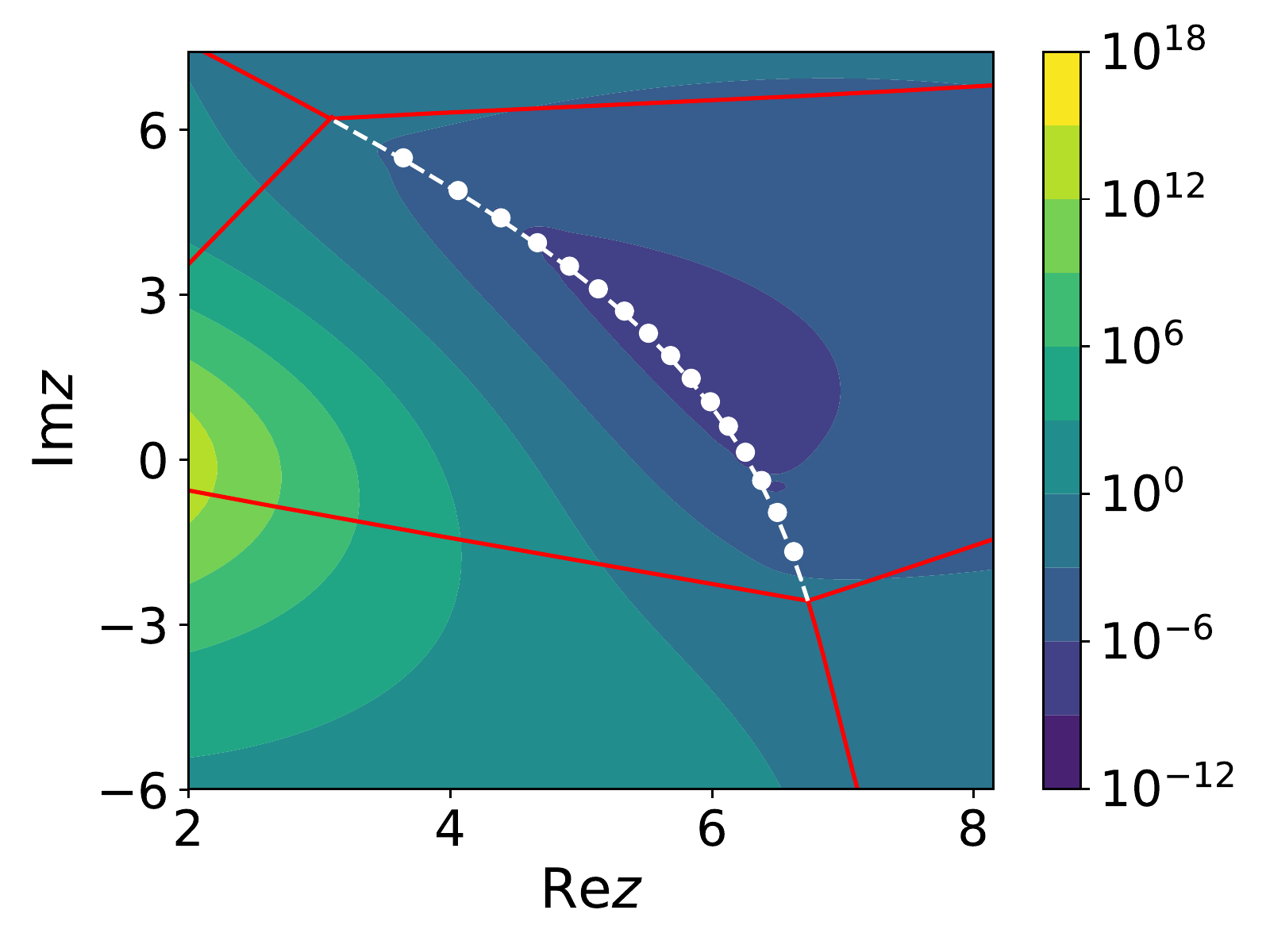}
	\fig{7.5cm}{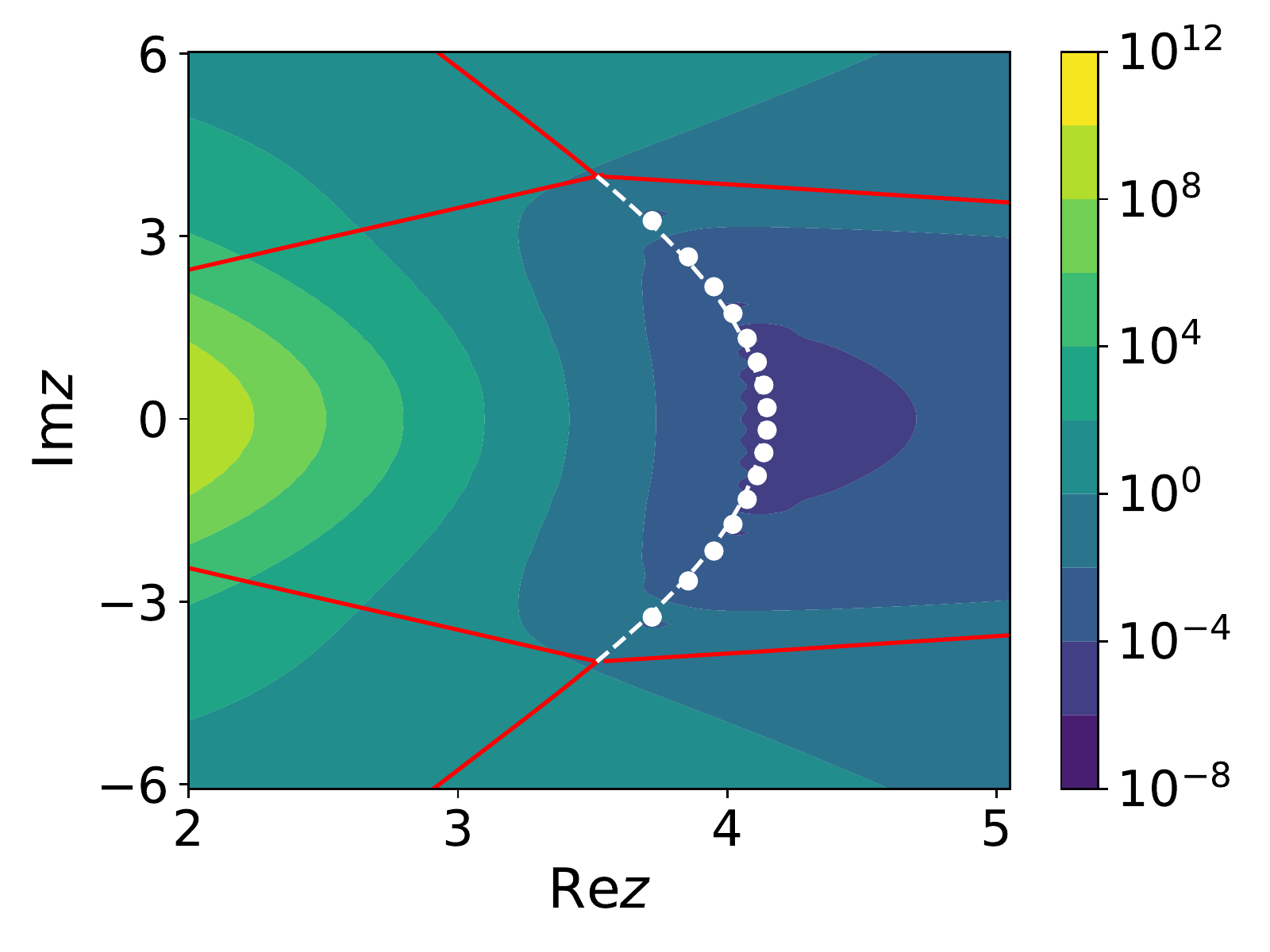}
	\caption{
	The value of $R_K$ at $K=16$ in the complex $z$-plane for $\lambda=1+1i$ and $\omega^2=1$ (left) and $\lambda=1$ and $\omega^2=-1$.
	Symbols and lines are same as Fig.~\ref{fig:stokes}.
    }
	\label{fig:RK appendix}
\end{figure}

We show behaviors of the reminder $R_K(z)$ defined by Eq.~\eqref{reminder} for a complex coupling constant and a double well potential $\omega^2 = -1$ in Fig.~\ref{fig:RK appendix}.
Since the Stokes and anti-Stokes lines are calculated in the complex $w$-plane, $\lambda$ and $\omega^2$ dependence enter through the mapping from $w$ to $z$ given in Eq.~\eqref{z to w}.
Again, we find a deep and flat valley around the anti-Stokes line similar to that shown in Fig.~\ref{fig:RK}.

%


\clearpage
\bibliographystyle{apsrev4-1}
\bibliography{opt.bib}

\end{document}